\documentclass[prl,twocolumn,floatfix,amsmath,nofootinbib,superscriptaddress,amssymb,preprintnumbers,floatfix]{revtex4-1}
\usepackage{amsmath,txfonts,longtable,booktabs,overpic,amssymb,bm,bbm,multirow,float,graphicx,color,dcolumn,subfigure,hyperref,orcidlink,tikz,fancyhdr}
\definecolor{blue}{RGB}{45,48,146}
\hypersetup{colorlinks,citecolor=blue,anchorcolor=red,menucolor=red, linkcolor=red,filecolor=red,runcolor=red,urlcolor=blue,frenchlinks=true}

\begin{document}

\title{Axial current as the origin of quantum intrinsic orbital angular momentum}

\author{Orkash Amat\,\orcidlink{0000-0003-3230-7587}}
\email{Correspondence Author. email: orkashamat@nju.edu.cn}
\affiliation{ School of Astronomy and Space Science, \href{https://ror.org/01rxvg760}{Nanjing University}, Nanjing 210023, China}

\author{Nurimangul Nurmamat\,\orcidlink{0000-0001-9227-3716}}
\affiliation{ School of Astronomy and Space Science, \href{https://ror.org/01rxvg760}{Nanjing University}, Nanjing 210023, China}

\author{Yong-Feng Huang \,\orcidlink{0000-0001-7199-2906}}
\email{Correspondence Author. email: hyf@nju.edu.cn}
\affiliation{ School of Astronomy and Space Science, \href{https://ror.org/01rxvg760}{Nanjing University}, Nanjing 210023, China}
\affiliation{ Key Laboratory of Modern Astronomy and Astrophysics (\href{https://ror.org/01rxvg760}{Nanjing University}), Ministry of Education, China}

\author{Cheng-Ming Li\,\orcidlink{0000-0002-9159-8129}}
\affiliation{ School of Physics and Microelectronics, \href{https://ror.org/04ypx8c21}{Zhengzhou University}, Zhengzhou 450001, China}

\author{Jin-Jun Geng\,\orcidlink{0000-0001-9648-7295}}
\affiliation{ \href{https://ror.org/02eb4t121}{Purple Mountain Observatory}, \href{https://ror.org/034t30j35}{Chinese Academy of Sciences}, Nanjing 210023, China}

\author{Chen-Ran Hu\,\orcidlink{0000-0002-5238-8997}}
\affiliation{ School of Astronomy and Space Science, \href{https://ror.org/01rxvg760}{Nanjing University}, Nanjing 210023, China}

\author{Ze-Cheng Zou\,\orcidlink{0000-0002-6189-8307}}
\affiliation{ School of Astronomy and Space Science, \href{https://ror.org/01rxvg760}{Nanjing University}, Nanjing 210023, China}

\author{Xiao-Fei Dong\,\orcidlink{0009-0000-0467-0050}}
\affiliation{ School of Astronomy and Space Science, \href{https://ror.org/01rxvg760}{Nanjing University}, Nanjing 210023, China}

\author{Chen Deng\,\orcidlink{0000-0002-2191-7286}}
\affiliation{ School of Astronomy and Space Science, \href{https://ror.org/01rxvg760}{Nanjing University}, Nanjing 210023, China}

\author{Fan Xu\,\orcidlink{0000-0001-7943-4685}}
\affiliation{ Department of Physics, \href{https://ror.org/05fsfvw79}{Anhui Normal University}, Wuhu, Anhui 241002, China}

\author{Xiao-li Zhang\,\orcidlink{0000-0002-3877-9289}}
\affiliation{ Department of Physics, \href{https://ror.org/01rxvg760}{Nanjing University}, Nanjing 210093, China}

\author{Chen Du\,\orcidlink{0009-0002-8460-1649}}
\affiliation{ School of Astronomy and Space Science, \href{https://ror.org/01rxvg760}{Nanjing University}, Nanjing 210023, China}

\begin{abstract}
We show that the axial current density is the physical origin (generator) of quantum intrinsic orbital angular momentum (IOAM). Without the axial current, the IOAM of particles vanishes. Broadly speaking, we argue that the spiral or interference characteristics of the axial current density determine the occurrence of nonlinear or tunneling effects in any spacetime-dependent quantum systems. Our findings offer a comprehensive theoretical framework that addresses the limitations of Keldysh's ionization theory and provides new insights into the angular momentum properties of quantum systems, particularly in tunneling-dominated regimes. Using Wigner function methods, fermionic generalized two-level model, and Berry phase simulations, we predict that IOAM effect can persist even in pure quantum tunneling processes. These results open the door for experimental verification of IOAM effects in future high-intensity QED experiments, such as those using X-ray free electron lasers.
\end{abstract}
\maketitle

\emph{Introduction.---} The fundamental concepts of the mechanical orbit angular momentum (often referred to extrinsic orbital angular momentum, EOAM)~\cite{newton1687}, spin angular momentum (SAM)~\cite{uhlenbeck1925} and  intrinsic orbital angular momentum (IOAM)~\cite{Allen:1992zz} play an indispensable role in the study and application of physics. Unlike EOAM, which is defined as $\mathbf{L}=\mathbf{r} \times \mathbf{P}$, IOAM describes quantum vortex particles with helical wavefronts characterized by a transverse phase factor ${\rm exp}\left(i\ell \phi\right)$ along the propagation direction~\cite{Jentschura:2010ap,PhysRevLett.107.174802,Bliokh:2015yhi,Padgett:17}. The IOAM has demonstrated significant potential in various fields, including strong-laser physics, nuclear physics, particle physics, and astrophysics~\cite{Bliokh:2017uvr,Ivanov:2022jzh,Session:2023hoq}. Recent efforts to address the transfer mechanisms of the IOAM have highlighted several critical advancements. These include the generation of vortex $\gamma$ photons and leptons through spin-to-orbital angular momentum transfer in nonlinear Compton scattering and nonlinear Breit-Wheeler processes~\cite{Ababekri:2022mob,Ababekri:2024cyd}, the study of helicity transfer mechanisms in the Schwinger effect via multiphoton pair production~\cite{Kohlfurst:2022edl}, and the development of quantum manipulation techniques~\cite{Lu:2023wrf,Jiang:2024fit}, dynamics of relativistic vortex electrons in strong background field~\cite{Ababekri:2024glc}.

Significant progress has been made in understanding the IOAM through both theoretical and experimental approaches. However, the physical nature, formation mechanisms, and origins of IOAM remain unresolved, making these areas subjects of active research. A central unresolved question is which physical quantity fundamentally determines IOAM formation. In the multiphoton-absorption regime, the IOAM magnitude is often captured by the Keldysh (ionization) parameter~\cite{Keldysh:1965ojf} or by effective-mass arguments~\cite{Kohlfurst:2013ura}; by contrast, cases involving tunneling, multiple concurrent mechanisms~\cite{Kohlfurst:2019mag}, or dynamical assistance~\cite{Schutzhold:2008pz} remain significantly more challenging.
To date, calculating the total IOAM generated within a system or spatial region from a quantum perspective remains a formidable challenge. Additionally, most studies on particles carrying IOAM have primarily focused on multiphoton processes. This raises a critical question: is it truly impossible to generate vortex particles carrying IOAM through tunneling processes? Moreover, although theoretical studies on the observation and calculation of particle OAM are presented in Ref.~\cite{Ji:1996ek}, it remains unclear whether this OAM belongs to the IOAM or EOAM. Exploring this possibility is an important direction for future research.
While Keldysh's ionization theory has successfully predicted the occurrence of multiphoton and tunneling processes in purely time- or space-dependent background fields~\cite{Dunne:2005sx,Dunne:2006st,Amat:2022uxq}, its applicability is significantly constrained in non-plane-wave, spacetime-dependent electromagnetic fields~\cite{Schneider:2014mla,Schneider:2018huk,Amat:2023vwv}. It is important to note that the Keldysh parameter, defined as $\gamma = {m \omega_0}/{e \varepsilon_0 E_{\text{cr}}}$, plays a crucial role in pair-creation studies~\cite{Dunne:2005sx}. Depending on its value, different production mechanisms can be identified: the multiphoton absorption regime corresponds to $\gamma \gg 1$, the tunneling regime to $\gamma \ll 1$, and an intermediate mixed regime to $\gamma \sim 1$~\cite{Kohlfurst:2019mag}. Furthermore, no comprehensive theoretical framework currently exists to describe the probability distribution of a particle's IOAM. In short, the general explanation and physical origins of IOAM remain open and unresolved questions, requiring further exploration.

In this Letter, we show that the IOAM originates from the axial current density. Without the axial current density, there would be no quantum IOAM of particles. The spiral (helical) or interference characteristics of the axial current density determine the occurrence of nonlinear or tunneling effects in the system. This result holds for any spacetime-dependent background field and addresses the shortcomings and limitations of Keldysh's ionization theory. By employing the Wigner function method, fermionic generalized two level model, and Berry phase simulations, we have explored potential future experiments on the Schwinger effect. Our findings reveal that the quantum IOAM effect persists even in pure quantum tunneling processes. This may provides new theoretical insights into the angular momentum properties of quantum systems.

\emph{Generalized quantum angular momentum.---}
To advance the understanding of the physical nature, formation mechanism, and origin of IOAM, we develop a generalized quantum angular momentum framework based on the Wigner function~\cite{Peskin:2018}. This approach allows us to move beyond the limitations of Keldysh's ionization theory. It thereby provides a more universal theoretical foundation for addressing these fundamental questions. The total quantum angular momentum of the quantum system in arbitrarily spacetime-dependent background fields can be written as
\begin{equation}\label{eq:1}
\mathcal{M} = \overbrace{\int d\Gamma \left(\mathbf{x} \times \mathbf{p}\right) \mathbbm{V}_0}^{{\rm EOAM}~of~particles}  + \overbrace{\int d\Gamma \frac{(-\mathbbm{A})}{2}}^{{\rm IOAM}~of~particles} + \overbrace{\int d^3x~ \mathbf{x} \times \left( \mathbf{E} \times \mathbf{B}\right)}^{{\rm EOAM} ~of~gauge~field},
\end{equation}
where the first, second, and third terms on the right-hand side of the above equation represent the EOAM of the particles, the IOAM of the particles, and the EOAM of the gauge field, respectively. The phase-space volume is defined as $d\Gamma=d^3xd^3p/(2\pi\hbar)^{3}$~\cite{Bialynicki-Birula:1991jwl,Hebenstreit:2011}, in which $\hbar$ is the reduced Planck constant. Here, $\mathbbm{V}_0$ and $\mathbbm{A}$ denote the charge and axial current densities, respectively\cite{Bialynicki-Birula:1991jwl,Sheng:2019ujr}. The key point is that the SAM and IOAM are distinct degrees of freedom of a particle~\cite{Bliokh:2015yhi,Ababekri:2022mob,Ababekri:2024cyd}. During multiphoton absorption, the SAM carried by the absorbed photons is transferred to the particle (electron) and manifests as IOAM, while the particle's SAM remains $\hbar/2$. Although in Ref.~\cite{Bialynicki-Birula:1991jwl} the quantity entering our definition of IOAM in Eq.~\eqref{eq:1} is described as a “spin density,” more recent studies interpret it as the axial (chiral) current density in Ref.~\cite{Sheng:2019ujr}. While it is related to spin density, it need not be directly tied to the spin of an individual particle or photon; more generally, it may encode multiphoton SAM transfer in nonlinear processes. Additionally, $\mathbf{x}$, $\mathbf{p}$, $\mathbf{E}$ and $\mathbf{B}$ are the vector position space, momentum, electric field and magnetic field, respectively.
If redefine that $\mathcal{L}_{\rm IOAM}=-\mathbbm{A}/2$ is as intrinsic orbital angular momentum probability density, $\mathcal{L}_{\rm IOAM}$ determines the occurrence of nonlinear effects or tunneling effects in any quantum systems. In this work, $\mathcal{L}_{\rm IOAM}$ provides a useful descriptor alongside the Keldysh parameter. It allows us to estimate the total IOAM of the quantum system and can help in interpreting possible signatures of quantum-vortex structures.

\emph{Generalized spin resolved fermionic quantum two level model.---}
To substantiate our results, we compute the topological charge (IOAM quantum number) $\ell$ of the produced particles and antiparticles, $\ell=\frac{\phi_{\rm Berry}}{2\pi}=\frac{1}{2\pi}\int_\mathcal{C} {\mathcal{A}}\cdot d\bf{q}$, where $\phi_{\rm Berry}$ is the Berry phase~\cite{Berry:1984jv,PhysRevA.61.032110}, ${\mathcal{A}} = \nabla_{\bf{q}}\left(\arg \left[c_{\bf{q}}^{(2)}(t)\right]\right)$. The term ``arg'' represents the argument of the complex
function $c_{\bf{q}}^{(2)}(t)$, i.e., $\arg
\left[c_{\bf{q}}^{(2)}(t)\right]= {\rm arctan}\left(
\Im\left[c_{\bf{q}}^{(2)}(t)\right]/\Re\left[c_{\bf{q}}^{(2)}(t)\right]
\right)$, referencing the foundational work of Berry~\cite{Berry:1984jv,PhysRevA.61.032110}. The $c_{\bf{q}}^{(2)}(t)$ can be obtained from the fermionic generalized two-level model  for studying the pair production under any time-dependent electric field in our previous work~\cite{Amat:2024nvg} as
\begin{eqnarray}\label{eq:2}
c_{\bf{q}}^{(2)}(t)=\sum_{\bf s} \sum_{\bf s'} c_{\bf{q},\bf{s},\bf{s}'}^{(2)}(t),
\end{eqnarray}
where
\begin{equation}\label{eq:3}
i\frac{d}{dt}\begin{bmatrix}
c_{\bf{q},\bf{s},\bf{s}'}^{(1)}(t)\\
c_{\bf{q},\bf{s},\bf{s}'}^{(2)}(t)
\end{bmatrix}
=
\begin{pmatrix} \omega_{\bf{q}}(t) & i\Omega_{\bf{q},\bf{s},\bf{s}'}(t) \cr -i\Omega^*_{\bf{q},\bf{s},\bf{s}'}(t) & -\omega_{\bf{q}}(t) \end{pmatrix}
\begin{bmatrix}
c_{\bf{q},\bf{s},\bf{s}'}^{(1)}(t)\\
c_{\bf{q},\bf{s},\bf{s}'}^{(2)}(t)
\end{bmatrix},
\end{equation}

\begin{equation}\label{eq:4}
\Omega_{\bf{q},\bf{s},\bf{s}'}(t)=\frac{u^\dagger_{\bf{q},\bf{s}'}(t) \dot{H}_{\bf{q}}(t) v_{\bf{q},\bf{s}}(t)}{2\omega_{\bf{q}}(t)},
\end{equation}
\begin{widetext}
\begin{align}\label{eq:5}
H_\mathbf{q}(t)=
\left(
\begin{array}{cccc}
m  & 0 &  p_z (t) &  (p_x (t)-ip_y (t)) \\
0 & m  &  (p_x (t)+ip_y (t)) & - p_z (t) \\
p_z (t) &  (p_x (t)-ip_y (t)) & -m & 0 \\
(p_x (t)+ip_y (t)) & - p_z (t) & 0 & -m  \\
\end{array}
\right),
\end{align}
\end{widetext}
and the initial conditions are $c_{\bf{q},\bf{s},\bf{s}'}^{(1)}(t_0)=1$ and $c_{\bf{q},\bf{s},\bf{s}'}^{(2)}(t_0)=0$. The kinetic momentum is $\mathbf{p}(t) = \mathbf{q} - e\,\mathbf{A}(t)$, where $\mathbf{q}$ is the canonical momentum. Here $\omega_{\mathbf{q}}(t)=\sqrt{m^2 + {[\mathbf{q} - e \mathbf{A}(t)]}^2}$ is single particle energy. The bispinors are chosen in the following form
\begin{eqnarray}
u_{\bf{q},s} &=& \sqrt{\frac{(p^0 +m)}{2p^0}}
\begin{pmatrix}
w_s\\
\frac{\boldsymbol{\sigma} \cdot \mathbf{p}(t)}{(p^0 +m)} w_s
\end{pmatrix} \,, \label{eq:6}\\
v_{\bf{q},s} &=& \sqrt{\frac{(p^0 +m)}{2p^0}}
\begin{pmatrix}
\frac{\boldsymbol{\sigma} \cdot \mathbf{p}(t)}{(p^0 +m)} w_{-s}\\
w_{-s}
\end{pmatrix} \,,\label{eq:7}
\end{eqnarray}
where $p^0 = p_0 = \sqrt{m^2 + \boldsymbol{p}^2(t)}$, and $w_{+1} = (1, \, 0)^\mathrm{t}$, $w_{-1} = (0, \, 1)^\mathrm{t}$ \,. The Pauli matrices are 
\begin{align}\label{eq:8}
\sigma_1=\left(
\begin{array}{cc}
0 & 1 \\
1 & 0 \\
\end{array}
\right), \sigma_2=\left(
\begin{array}{cc}
0 & -i \\
i & 0 \\
\end{array}
\right), \sigma_3=\left(
\begin{array}{cc}
1 & 0 \\
0 & -1 \\
\end{array}
\right).
\end{align}
    
The momentum distribution for different spin of electron and positron can be calculated by using the coefficient $c_{\bf{q},\bf{s},\bf{s}'}^{(2)}$ at $t=+\infty$ as
\begin{eqnarray}\label{eq:9}
f_{\bf{q}}^{\bf{s}\bf{s}'}(+\infty)=2|c_{\bf{q},\bf{s},\bf{s}'}^{(2)}(t=+\infty)|^2.
\end{eqnarray}
The electron and positron  momentum distributions are
\begin{equation}\label{eq:10}
f_{\bf{q}}^{\bf{s}'}(+\infty)=\sum_{\bf s} f_{\bf{q}}^{\bf{s}\bf{s}'}(+\infty),\ \ f_{\bf{q}}^{\bf{s}}(+\infty)=\sum_{\bf s'} f_{\bf{q}}^{\bf{s}\bf{s}'}(+\infty).
\end{equation}
The total momentum distribution is
\begin{eqnarray}\label{eq:11}
f_{\bf{q}}(+\infty)=\sum_{\bf s} \sum_{\bf s'} f_{\bf{q}}^{\bf{s}\bf{s}'}(+\infty).
\end{eqnarray}

We will then conduct a systematic comparison with the theoretical predictions associated with the $\mathcal{L}_{\rm IOAM}$.

\emph{Dirac-Heisenberg-Wigner formalism.---}
The starting point is the Dirac equation expressed as:
\begin{equation}\label{E1}
\left( i\gamma^{\mu}\partial_\mu- e \gamma^{\mu} A_\mu(\mathbf{x},t)-m\right)\Psi(\mathbf{x},t)=0,
\end{equation}
where the four-gradient is denoted by $\partial_\mu=(\partial_t, \bm{\nabla})$. The $e$ and $m$ denote the electron charge and mass respectively. The four-potential of the electromagnetic field is $A_\mu(\mathbf{x},t)=(\varphi, -\mathbf{A})$.

Eq.~(\ref{E1}) can equivalently be expressed as
\begin{equation}\label{E2}
i\frac{\partial}{\partial t}\Psi(\mathbf{x},t)=\mathcal{H}(\mathbf{x},t)\Psi(\mathbf{x},t),
\end{equation}
with the time-dependent Hamiltonian given by
\begin{equation}\label{E3}
\mathcal{H}(\mathbf{x},t)=\bm{\alpha}\cdot\big[\mathbf{q}-e \mathbf{A}(\mathbf{x},t)\big]+\beta mc^2+e\varphi(\mathbf{x},t),
\end{equation}
where $\mathbf{q}=-i\bm{\nabla}$ is the canonical momentum operator, $\bm{\alpha}=\gamma^0\bm{\gamma}=\left(
           \begin{array}{cc}
             0 & \bm{\sigma} \\
             \bm{\sigma} & 0 \\
           \end{array}
         \right)$, and $\beta=\gamma^0$.

The equal-time density operator for spinor QED, defined as the commutator of two field operators in the Heisenberg picture, takes the form \cite{Hebenstreit:2011}
\begin{widetext}
\begin{align}\label{E4}
\hat{\mathcal{C}}_{spinor}(\mathbf{x}_1,\mathbf{x}_2,t)\equiv \exp\Big(-ie\int_{\mathbf{x}_2}^{\mathbf{x}_1} d\mathbf{x}'\cdot\mathbf{A}(\mathbf{x}',t)\Big) \left[\bar{\Psi} (\mathbf{x}_1,t),{\Psi}(\mathbf{x}_2,t) \right].
\end{align}
\end{widetext}
Introducing the center-of-mass coordinate $\mathbf{x}=(\mathbf{x}_1+\mathbf{x}_2)/2$ and the relative coordinate $\mathbf{r}=\mathbf{x}_1-\mathbf{x}_2$, the density operator becomes
\begin{widetext}
\begin{align}\label{E5}
\hat{\mathcal{C}}_{spinor}(\mathbf{x},\mathbf{r},t)=\exp\Big(-ie\int_{-1/2}^{1/2}d\lambda\,\mathbf{r}\cdot\mathbf{A}(\mathbf{x}+\lambda\mathbf{r},t)\Big) \left[\bar{\Psi} \Big(\mathbf{x}+\frac{\mathbf{r}}{2},t\Big),\Psi \Big(\mathbf{x}-\frac{\mathbf{r}}{2},t\Big)\right].
\end{align}
\end{widetext}
The Wigner function is then defined as the vacuum expectation value of the Wigner operator, which corresponds to the Fourier transform of the equal-time commutator $\mathcal{C}_{spinor}(\mathbf{x},\mathbf{r},t)$ with respect to the relative coordinate $\mathbf{r}$:
\begin{eqnarray}\label{E6}
\mathcal{W}_{spinor}(\mathbf{x},\mathbf{p},t)=\frac{1}{2}\int{d^3 \mathbf{r}\,
\langle\mathrm{vac}|\hat{\mathcal{C}}_{spinor}(\mathbf{x},\mathbf{r},t)|\mathrm{vac}\rangle e^{-i\mathbf{p}\cdot\mathbf{r}}}.
\end{eqnarray}
The exponential factor in the integrand of Eq.~(\ref{E6}), known as the Wilson line factor, ensures gauge invariance. The integration path is chosen along a straight line, introducing a well-defined kinetic momentum $\mathbf{p}$. In this context, the Hartree approximation is applied, treating the electromagnetic field as a classical (C-number) rather than a quantum (Q-number) field.

By differentiating Eq.~(\ref{E6}) with respect to time and using Eq.~(\ref{E2}) with $\mathbf{A}(t)$ generalized to $\mathbf{A}(\mathbf{x},t)$, one derives the equation of motion for the Wigner function \cite{Hebenstreit:2011}:
\begin{widetext}
\begin{equation}\label{E7}
  D_t\mathcal{W}_{spinor}=-\frac{1}{2}\mathbf{D}\cdot\left[\gamma^0\bm{\gamma},\mathcal{W}_{spinor}\right]
  \!-i\mathbf{\Pi}\cdot\left\{\gamma^0\bm{\gamma},\mathcal{W}_{spinor}\right\}
  \!-im\left[\gamma^0,\mathcal{W}_{spinor}\right]
  ,
\end{equation}
\end{widetext}
where $D_t$, $\mathbf{D}$, and $\mathbf{\Pi}$ represent the pseudo-differential operators
\begin{alignat}{6}
  &D_t&\ =\ &\ \ \partial_t \ &\ + \ & e  \int_{-1/2}^{1/2}{d\xi\,\mathbf{E}(\mathbf{x}+i\xi\partial_\mathbf{p}},t)\cdot\partial_\mathbf{p} \ , \nonumber
  \\
  \label{E8}
  &\mathbf{D}&\ =\ &\ \bm{\nabla} \ & + \ & e  \int_{-1/2}^{1/2}{d\xi\,\mathbf{B}(\mathbf{x}+i\xi\partial_\mathbf{p},t)
  \times\partial_\mathbf{p}} \ ,
  \\
  &\mathbf{\Pi}&\ =\ &\ \ \mathbf{p}\ &\ - \ & ie  \int_{-1/2}^{1/2}{d\xi\,\xi\,\mathbf{B}(\mathbf{x}+i\xi\partial_\mathbf{p},t)
  \times\partial_\mathbf{p}} \ . \nonumber
\end{alignat}

The Wigner function $\mathcal{W}_{spinor}(\mathbf{x},\mathbf{p},t)$ can be decomposed into a complete basis set $\{\mathbbm{1},\gamma_5,\gamma^\mu,\gamma^\mu\gamma_5,\sigma^{\mu\nu}=\tfrac{i}{2}[\gamma^\mu,\gamma^\nu]\}$ with $16$ irreducible components (DHW functions): scalar $\mathbbm{s}(\mathbf{x},\mathbf{p},t)$ (mass density), pseudoscalar $\mathbbm{p}(\mathbf{x},\mathbf{p},t)$, vector $\mathbbm{v}_\mu(\mathbf{x},\mathbf{p},t)$ (${\mathbbm v}_0$ and ${\vec {\mathbbm v}}$ represent charge and current density), axialvector $\mathbbm{a}_\mu(\mathbf{x},\mathbf{p},t)$ (spin polarization density), and tensor $\mathbbm{t}_{\mu\nu}(\mathbf{x},\mathbf{p},t)$ (electric and magnetic dipole moments, $i,j=1,2,3$) \cite{Sheng:2019ujr}:
\begin{equation}\label{E9}
  \mathcal{W}_{spinor}(\mathbf{x},\mathbf{p},t)=\frac{1}{4}\left(\mathbbm{1}\mathbbm{s}+i\gamma_5\mathbbm{p}
+\gamma^\mu\mathbbm{v}_\mu+\gamma^\mu\gamma_5\mathbbm{a}_\mu+\sigma^{\mu\nu}\mathbbm{t}_{\mu\nu}\right).
\end{equation}

The vacuum initial conditions are chosen as
\begin{equation}\label{E10}
{\mathbbm s}_{\rm vac} = \frac{-2m}{\sqrt{{\mathbf p}^2+m^2}} \, ,
\quad  {\mathbbm v}_{i,{\rm vac}} = \frac{-2{ p_i} }{\sqrt{{\mathbf p}^2+m^2}}, \,
\end{equation}
where the subscript `$\rm vac$' indicates vacuum, and $i=1,2,3$ corresponds to the $x$, $y$, and $z$ axes.

In general, the equations governing the Wigner coefficients are integro-differential, whose solutions are numerically demanding due to their inherent non-local structure, see e.g. \cite{Hebenstreit:2011}.

For a spatially homogeneous electric field, the system reduces to a set of ordinary differential equations \cite{Blinne:2015zpa}. In this case, only ten of the sixteen coefficients remain nonzero:
\begin{equation}\label{E11}
{\mathbbm w} = ( {\mathbbm s},{\mathbbm v}_i,{\mathbbm a}_i,{\mathbbm t}_i)
\, , \quad  {\mathbbm t}_i := {\mathbbm t}_{0i} -   {\mathbbm t}_{i0}  \, .
\end{equation}
The kinetic momentum ${\mathbf p}$ relates to the canonical momentum ${\mathbf q}$ by
\begin{equation}\label{E12}
{\mathbf p}(t) = {\mathbf q} - e {\mathbf A} (t),
\end{equation}
which depends explicitly on time. The scalar Wigner coefficient is then connected to the one-particle distribution function $f({\mathbf q},t)$, itself related to the phase-space energy density:
\begin{equation}\label{E13}
\varepsilon = m {\mathbbm s} + p_i {\mathbbm v}_i.
\end{equation}
Hence, the momentum distribution reads
\begin{equation}\label{E14}
f({\mathbf q},t) = \frac 1 {2 \omega(\mathbf{q},t)} (\varepsilon - \varepsilon_{vac} ),
\end{equation}
with $\omega(\mathbf{q},t)= \sqrt{{\mathbf p}^2(t)+m^2}=
\sqrt{m^{2}+(\mathbf{q}-e\mathbf{A}(t))^{2}}$ as the single-particle energy.

It is convenient to redefine the following quantities:
\begin{align}\label{E15}
&{\mathbbm{S}} (\mathbf{p}(t),t):= (1-f({\mathbf q},t)) {\mathbbm s}_{vac}-\mathbf{p}(t) \cdot v_i (\mathbf{q},t),\\  \label{E16}
&\mathbbm{V}_i (\mathbf{q},t):= {\mathbbm v}_i (\mathbf{p}(t),t) -(1-f({\mathbf q},t)) {\mathbbm v}_{i,vac} (\mathbf{p}(t),t),\\  \label{E17}
&\mathbbm{A}_i(\mathbf{q},t):=\mathbbm{a}_i(\mathbf{q},t),\\ \label{E18}
&\mathbbm{T}_i(\mathbf{q},t):=\mathbbm{t}_i(\mathbf{q},t).
\end{align}

The single-particle momentum distribution $f(\mathbf{q},t)$ is obtained by solving the following ten coupled ODEs:
\begin{widetext}
\begin{equation}\label{E19}
\begin{array}{l}
\displaystyle
\dot{f}=\frac{e}{2\omega} \, \,  \mathbf{E}\cdot \mathbbm{V},\\[2mm]
\displaystyle
\dot{\mathbbm{V}}=\frac{2}{\omega^{3}}
\left( (e\mathbf{E}\cdot \mathbf{p})\mathbf{p}-e\omega^{2}\mathbf{E}\right) (f-1)
-\frac{(e\mathbf{E}\cdot \mathbbm{V})\mathbf{p}}{\omega^{2}}
-2\mathbf{p}\times \mathbbm{A} -m \mathbbm{T},\\[2mm]
\displaystyle
\dot{\mathbbm{A}}=-2\mathbf{p}\times \mathbbm{V},\\
\displaystyle
\dot{\mathbbm{T}}=\frac{2}{m}[m^{2}\mathbbm{V}+(\mathbf{p}\cdot \mathbbm{V})\mathbf{p}].
\end{array}
\end{equation}
\end{widetext}
The initial conditions are set as $f(\mathbf{p},-\infty)=\mathbbm{V}(\mathbf{p},-\infty)= \mathbbm{A}(\mathbf{p},-\infty)=\mathbbm{T}(\mathbf{p},-\infty)=0$ to carry out the calculations. Finally, the pair number density is given by integrating $f(\mathbf{q},t)$ over all momenta at asymptotically late times $t\to +\infty$:
\begin{equation}\label{E20}
  n(t\rightarrow\infty)= \lim_{t\to +\infty}\int\frac{d^{3}q}{(2\pi)^ 3}f(\mathbf{q},t) \, .
\end{equation}

\emph{Setup.---}
To validate the aforementioned perspective, we propose conducting theoretical research on potential future experiments related to the Schwinger effect~\cite{Ringwald:2001ib,Alkofer:2001ik,Heinzl:2009bmy,Fedotov:2022ely}. Our idealized experiment involves a representative setup using a circularly polarized (CP) electric field, which is generated by two counter-propagating short laser pulses. Notably, the magnetic fields of the two laser pulses cancel out in the standing wave, resulting in an electric field that is approximately spatially homogeneous throughout the entire interaction region. The electric field can be expressed in a time-dependent form as ${\bf E}\left(t\right)=\varepsilon_{0} E_{cr} d(t) \left({\rm cos}(\omega_{0} t),~{\rm sin}(\omega_{0} t ),~0 \right)$, $d(t) ={\rm sin}^{4}\left(\frac{\omega_{0} t}{2N}\right)$ for $0<\omega_{0}~t<2\pi N$, where $\varepsilon_{0}$ is the peak field strength, $E_{cr}$ denotes the Schwinger critical field strength, $\varphi=\omega_{0} t$ is the time-dependent phase of background field, $\omega_{0}$ and $N$ are the frequency and cycle number of the individual electric field, respectively. The pulse duration $\tau$ is given by $\tau=2\pi N/\omega_0=N \lambda/c$, where $\lambda$ is the wavelength and $c$ is the speed of light. In this latter we use the natural units ($\hbar=c=1$), and express all quantities in terms of the electron mass $m$.

The choice of this scheme is motivated by two primary reasons. First, the particle charge density $\mathbbm{v}_0$ generated in such a electric field is zero~\cite{Amat:2023vwv}. Second, the absence of a magnetic field eliminates the EOAM contribution from the gauge field. This ensures that IOAM can be investigated precisely without any interference from EOAM effects, see Eq.~\eqref{eq:1}.

\emph{Results.---}
\begin{figure*}[!t]\centering
\includegraphics[width=0.265\textwidth]{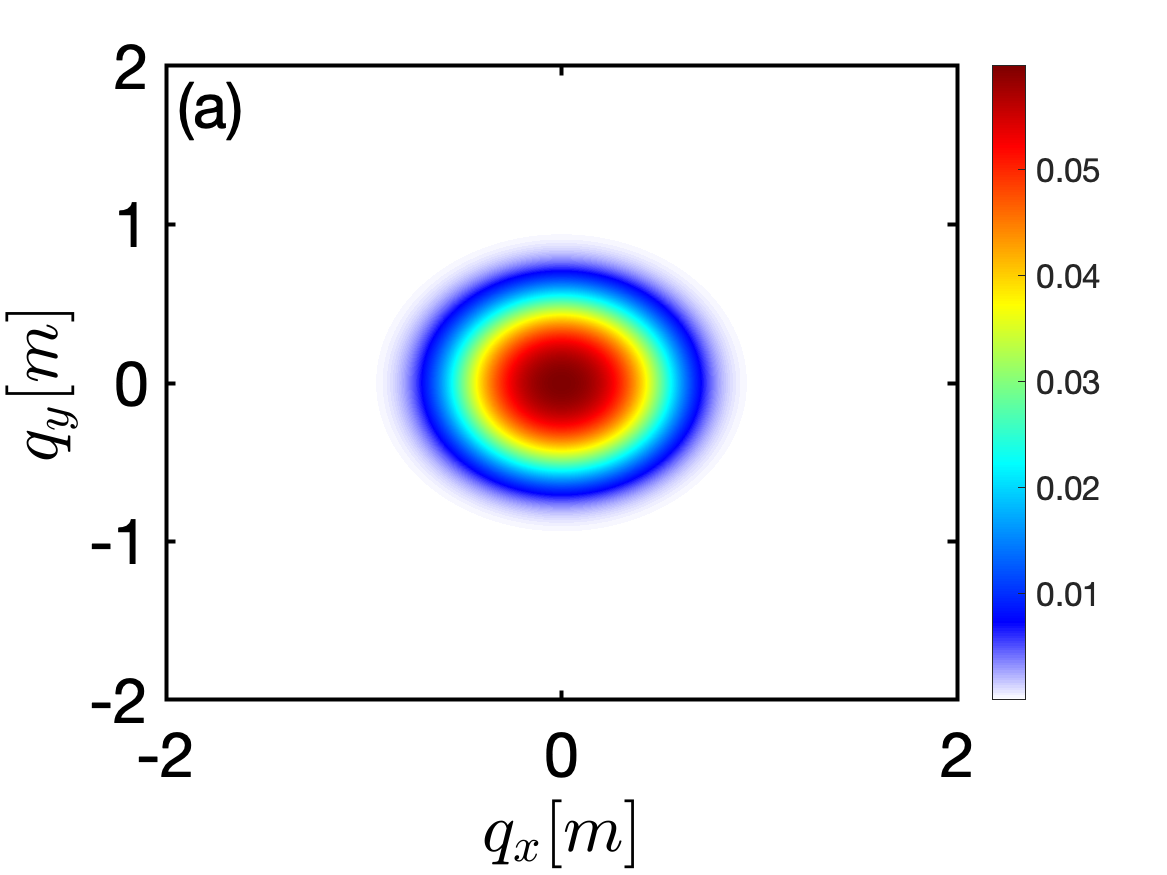}
\includegraphics[width=0.265\textwidth]{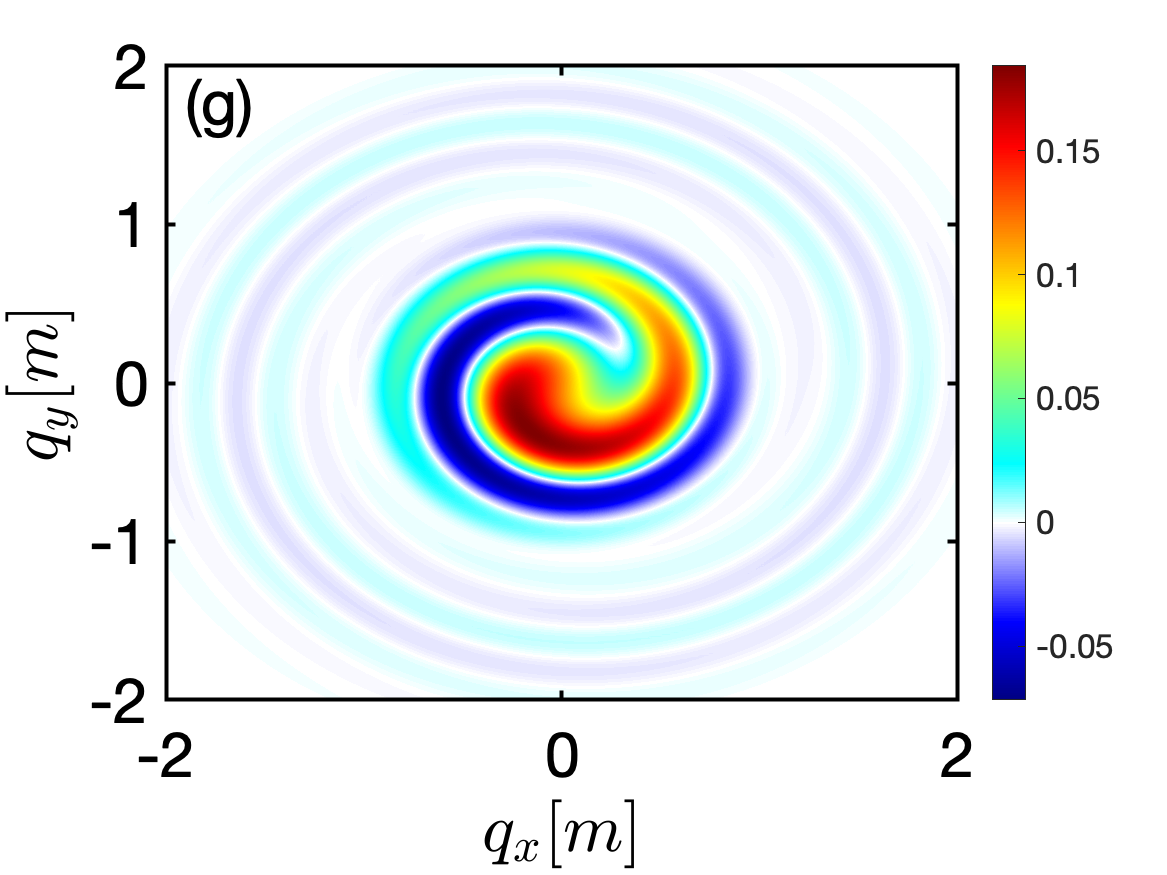}
\includegraphics[width=0.265\textwidth]{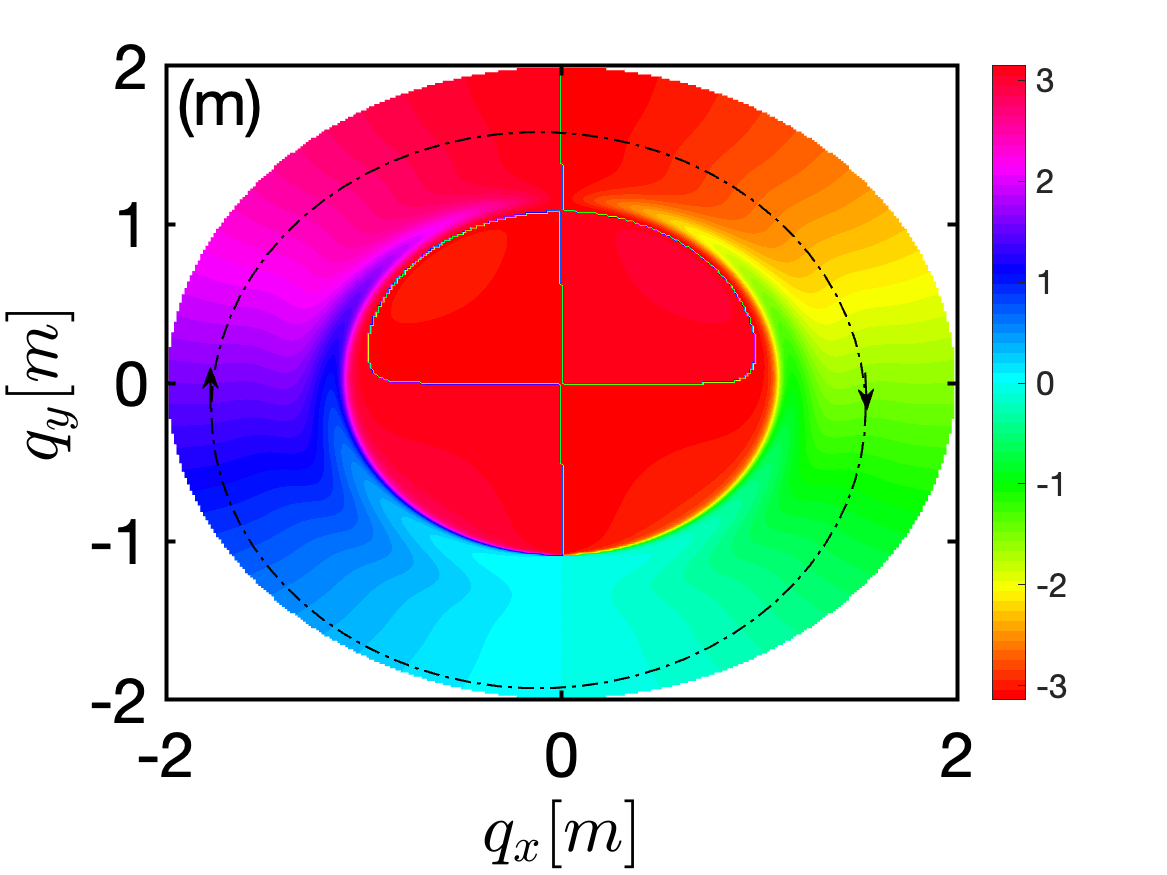}
\includegraphics[width=0.265\textwidth]{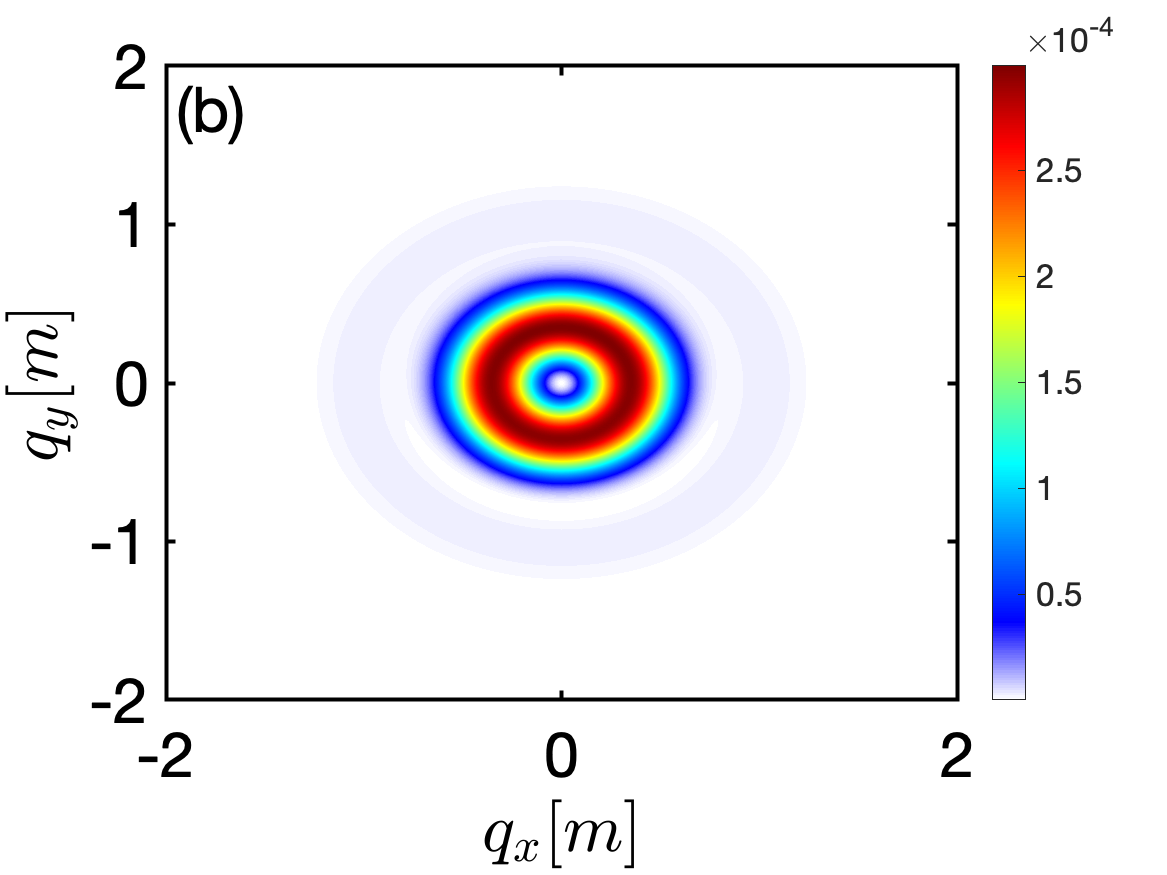}
\includegraphics[width=0.265\textwidth]{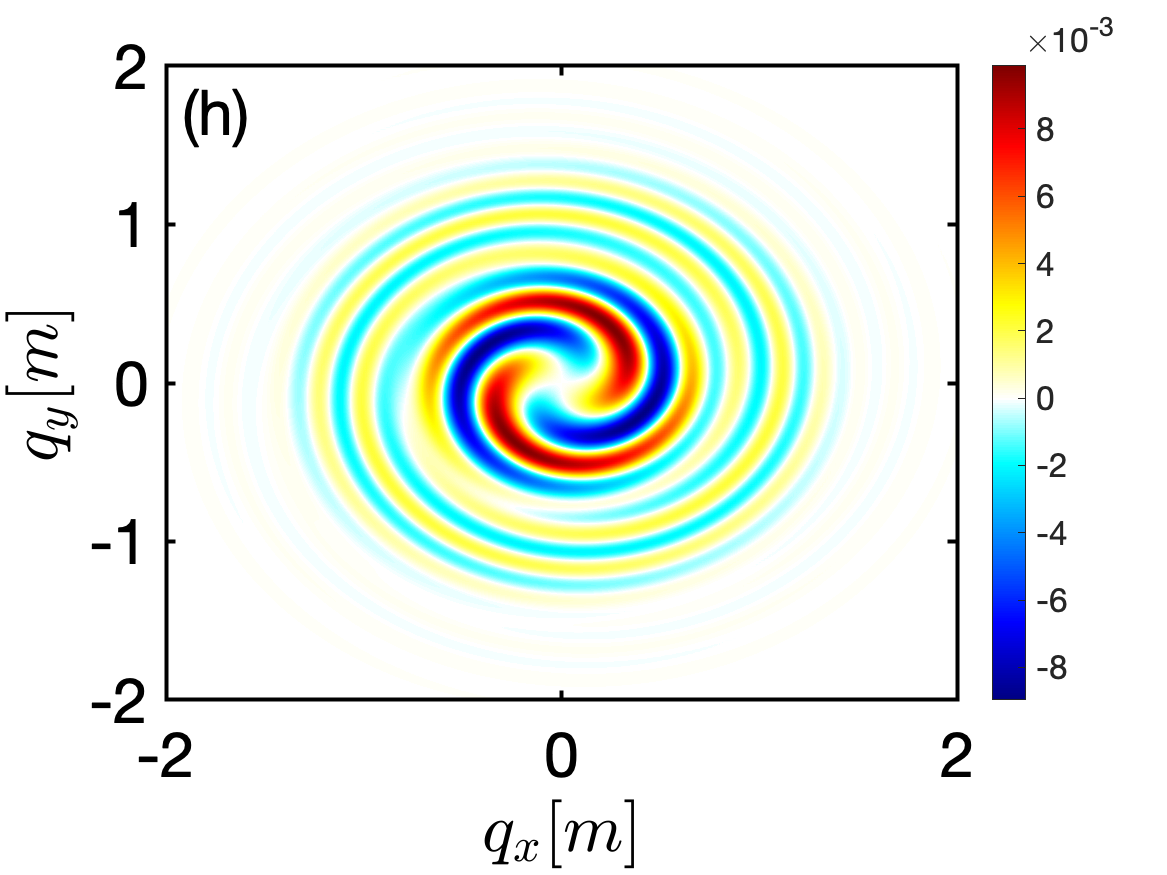}
\includegraphics[width=0.265\textwidth]{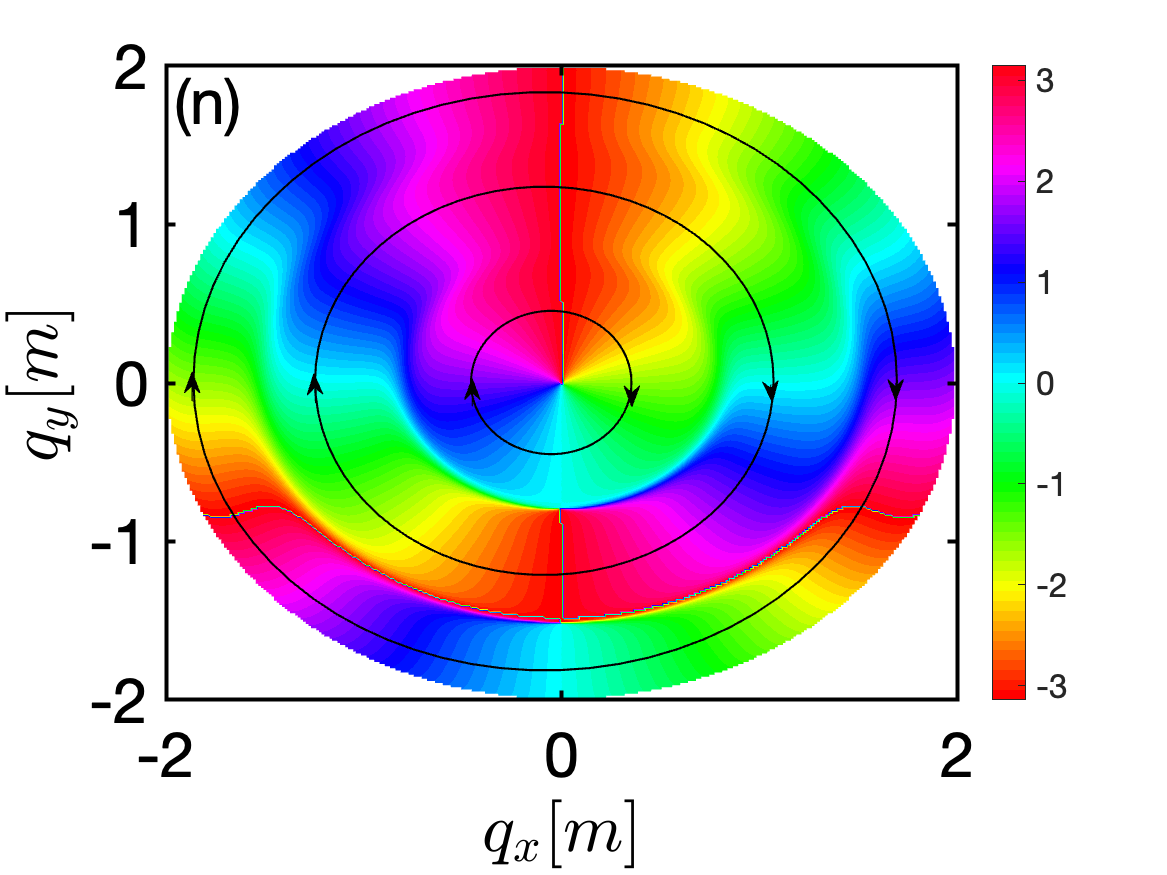}
\includegraphics[width=0.265\textwidth]{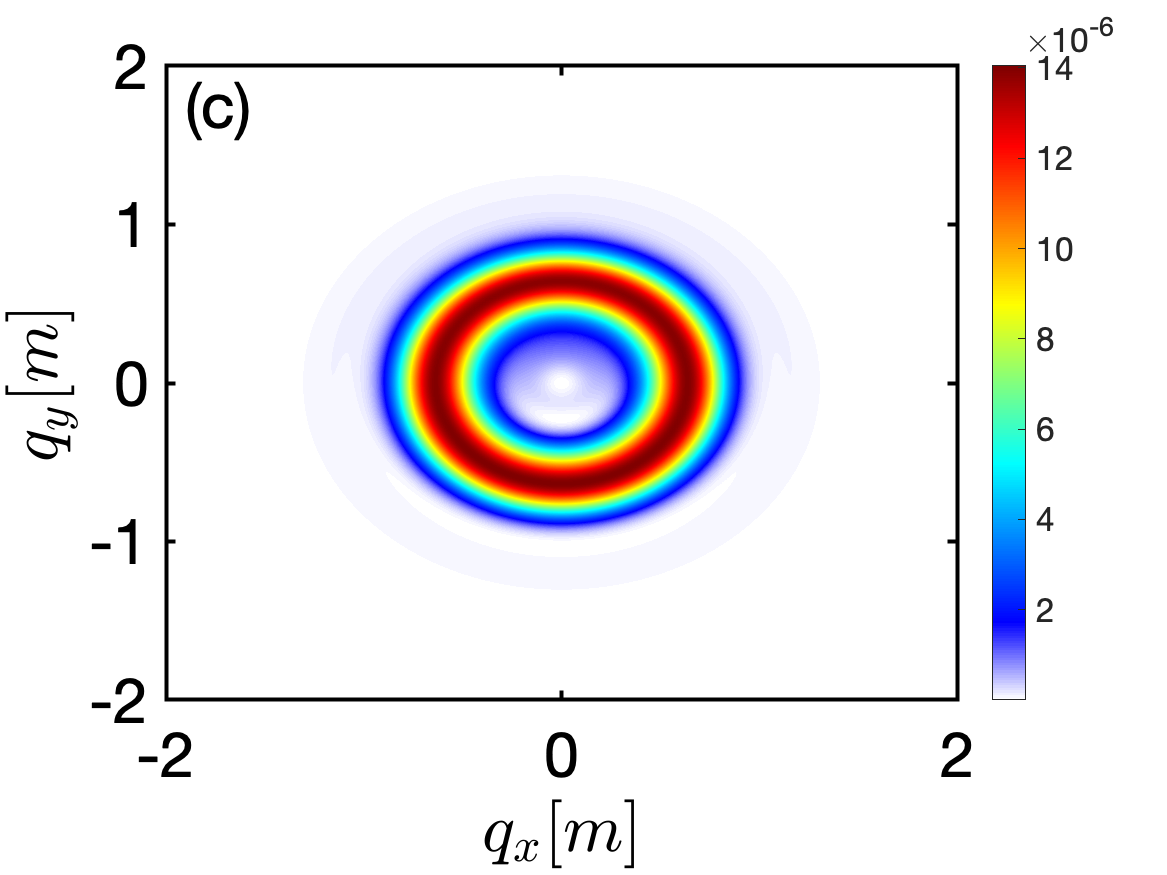}
\includegraphics[width=0.265\textwidth]{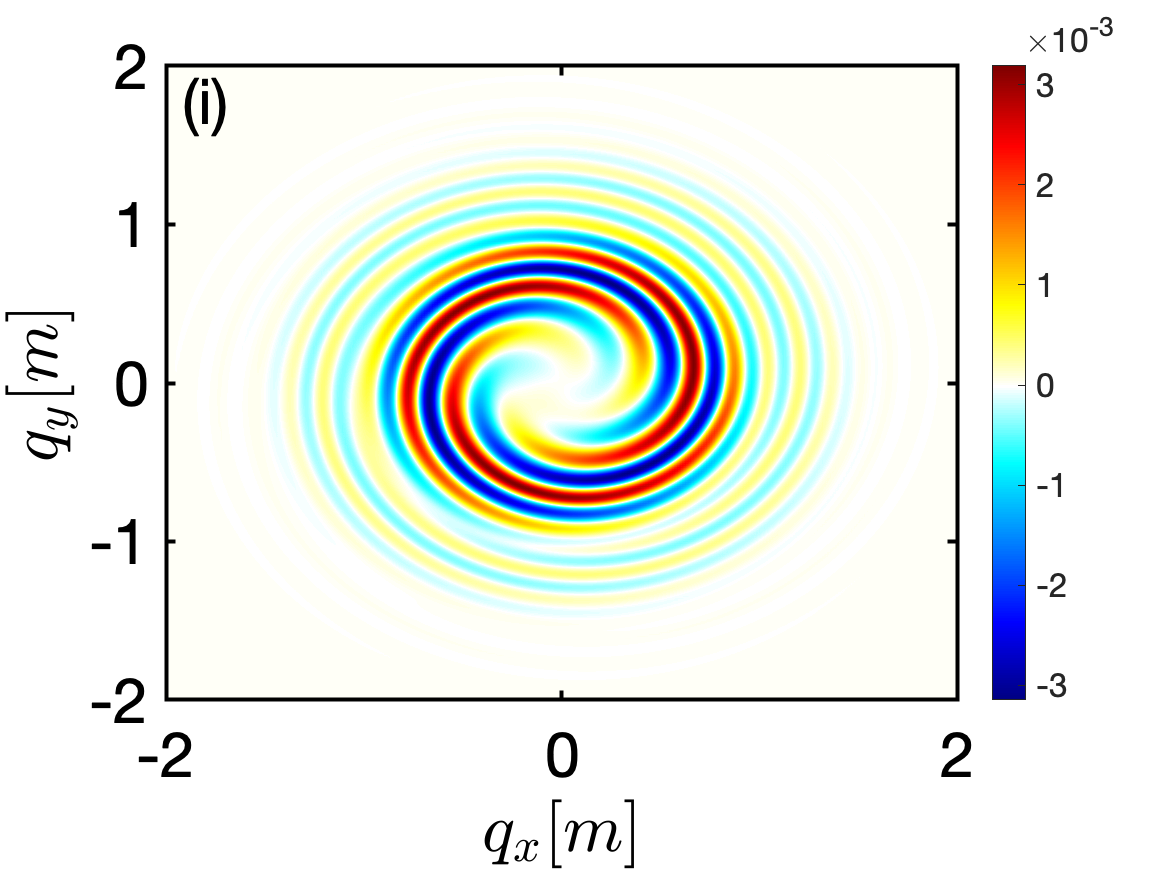}
\includegraphics[width=0.265\textwidth]{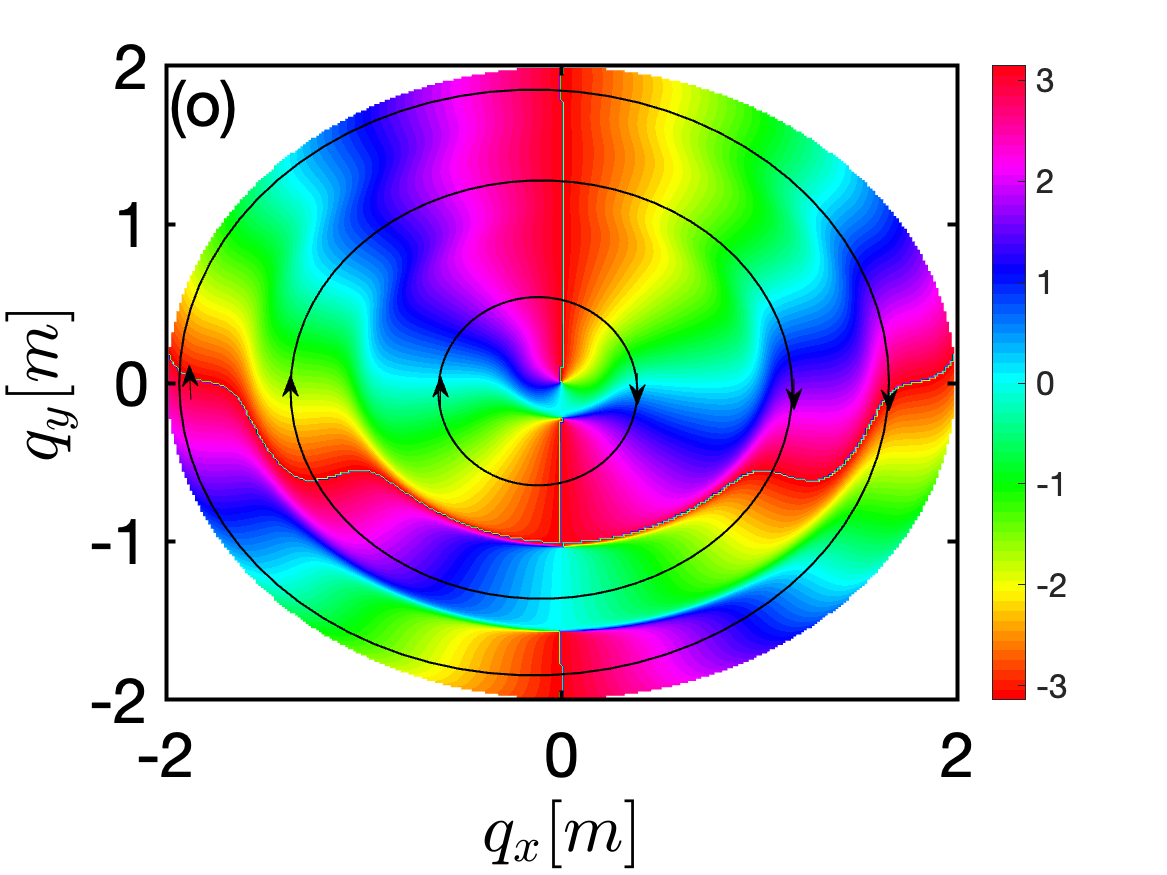}
\includegraphics[width=0.265\textwidth]{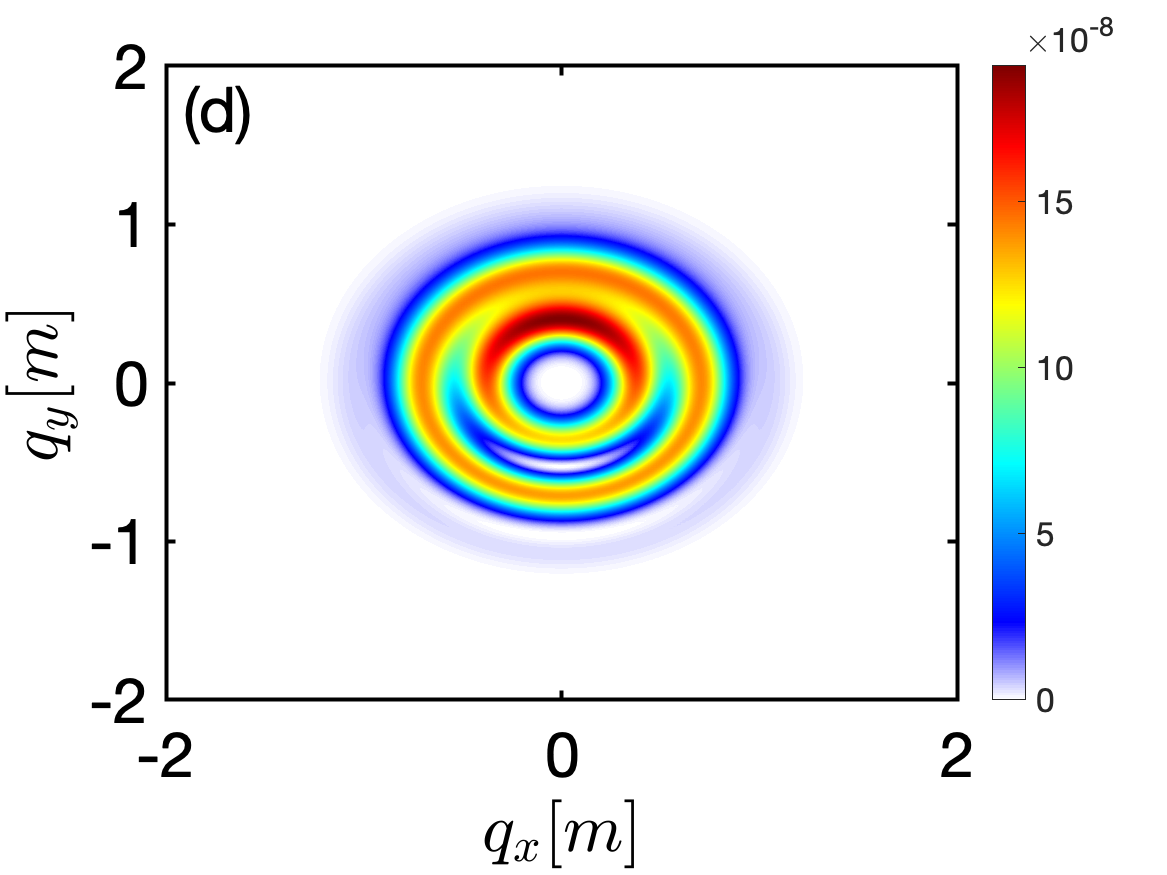}
\includegraphics[width=0.265\textwidth]{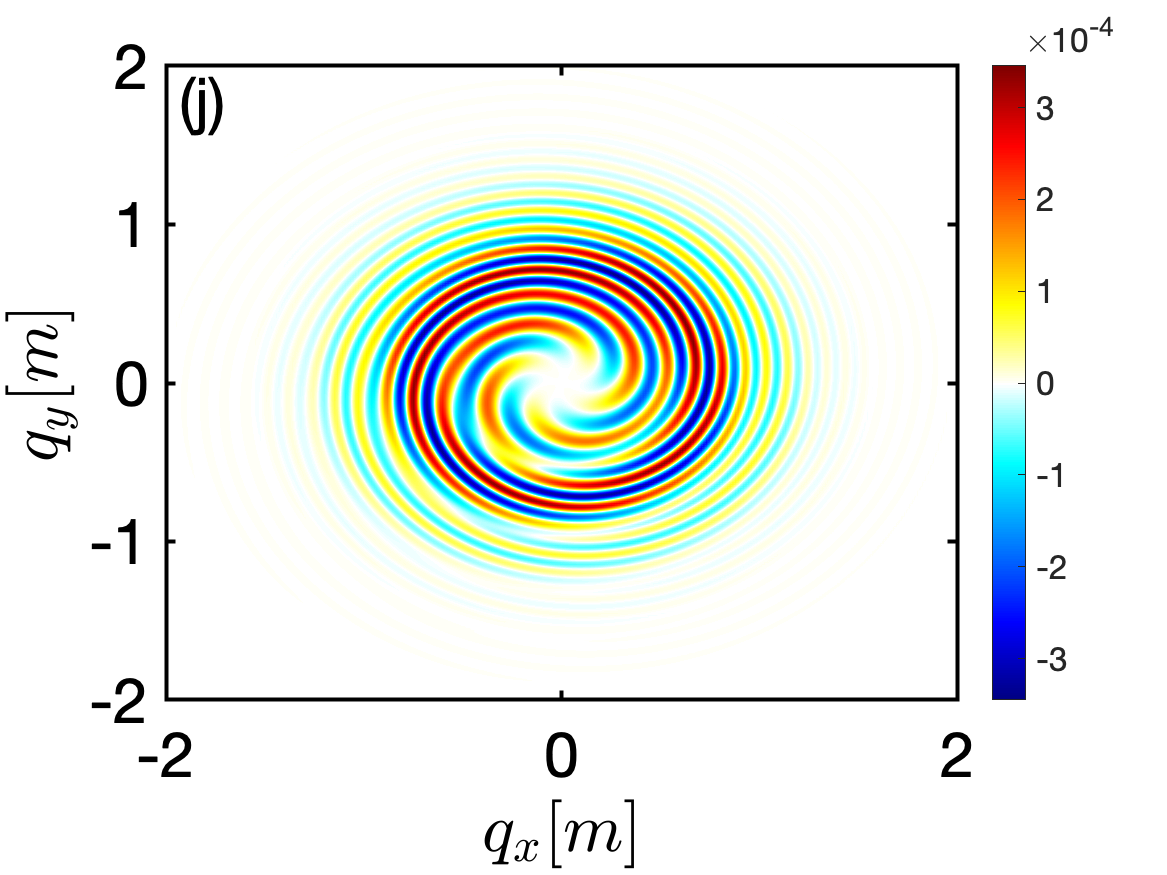}
\includegraphics[width=0.265\textwidth]{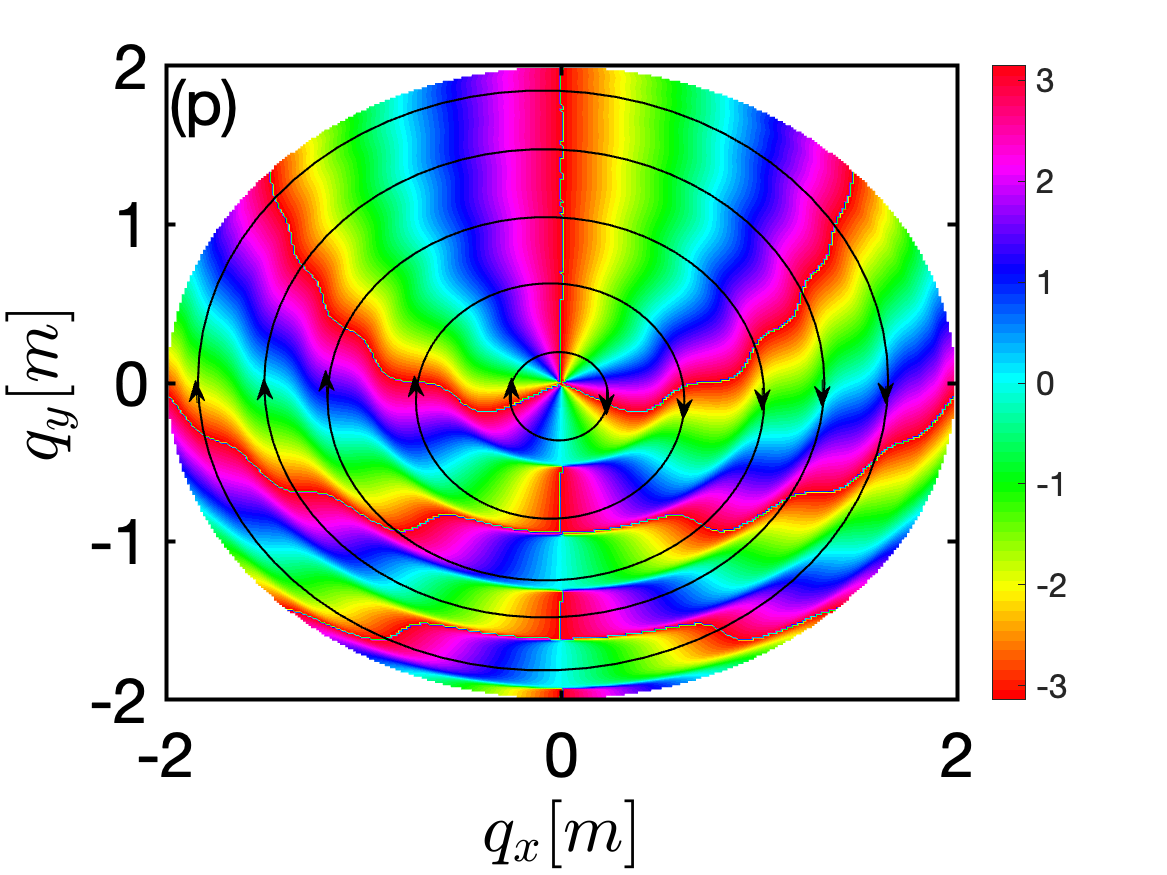}
\includegraphics[width=0.265\textwidth]{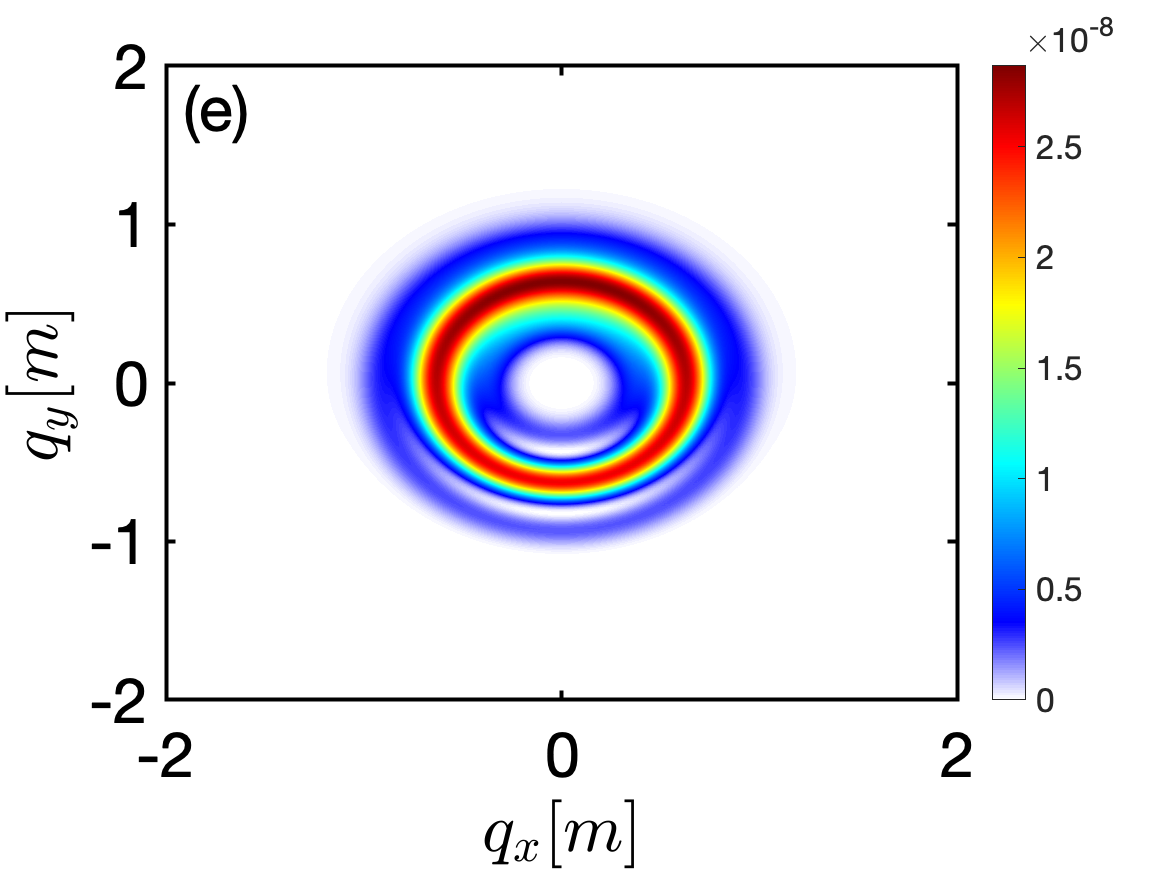}
\includegraphics[width=0.265\textwidth]{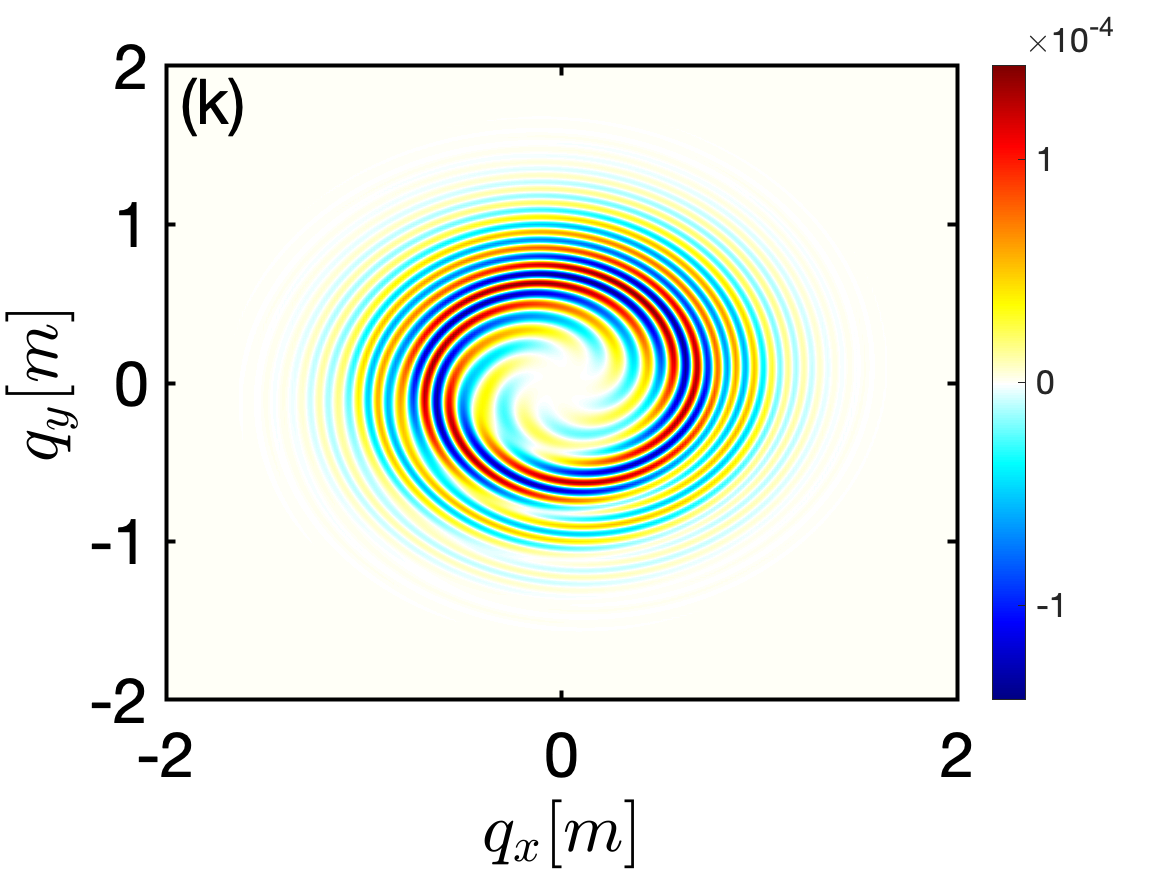}
\includegraphics[width=0.265\textwidth]{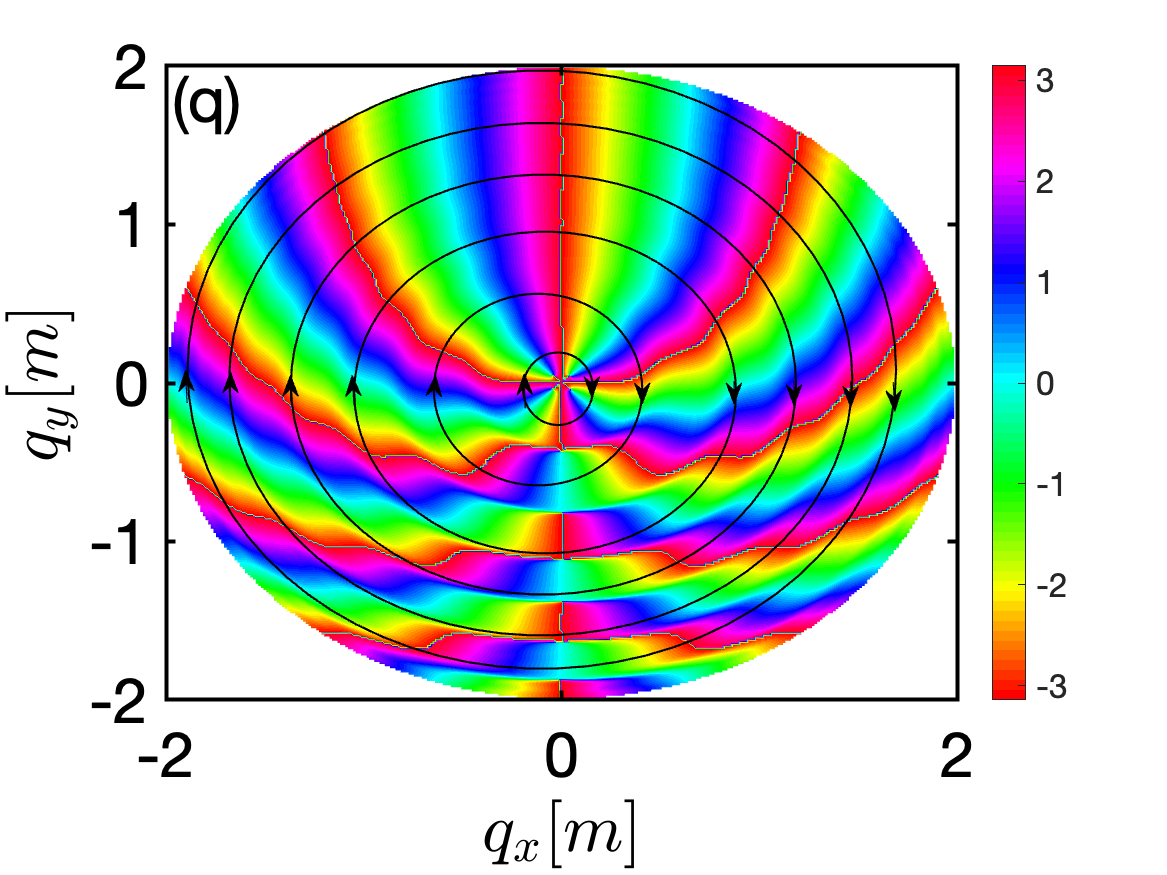}
\includegraphics[width=0.265\textwidth]{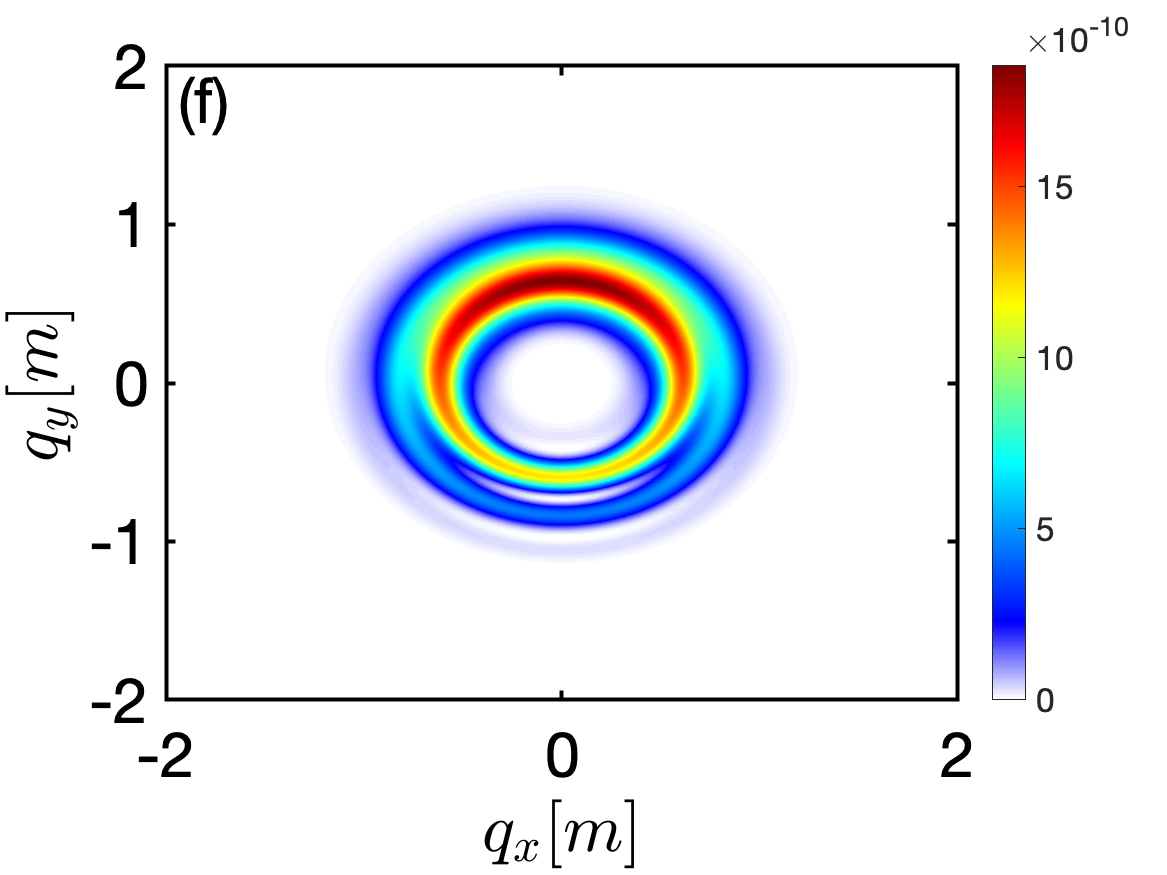}
\includegraphics[width=0.265\textwidth]{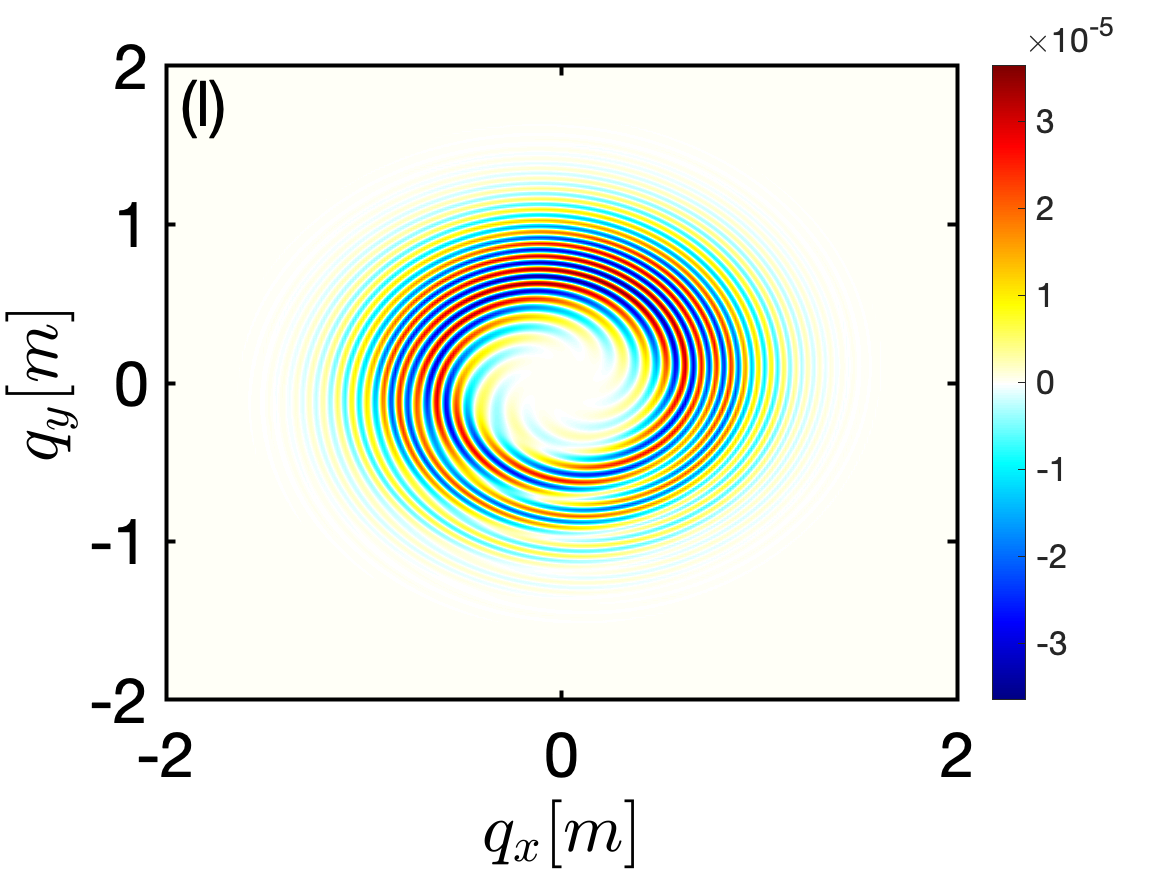}
\includegraphics[width=0.265\textwidth]{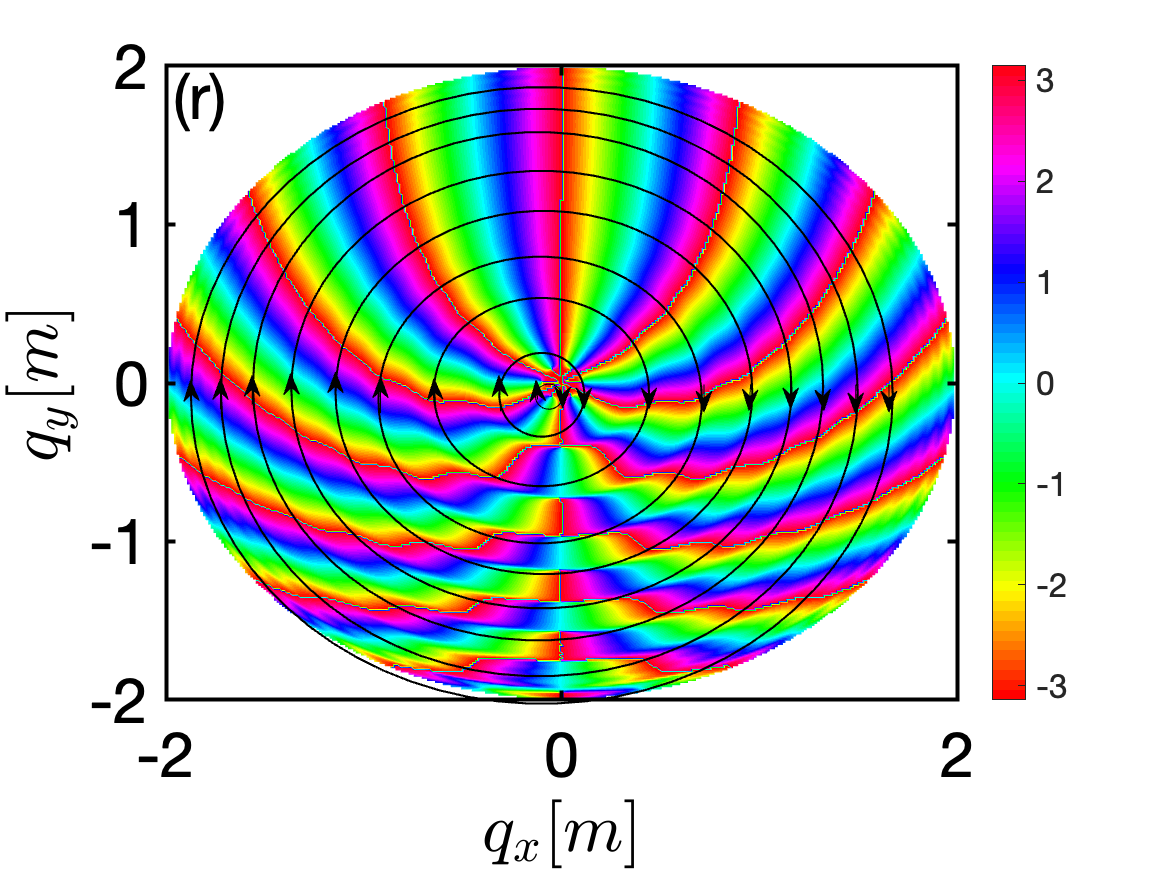}
\caption{Electron momentum distribution function $f_{\mathbf{q}}(+\infty)$ (a--f), intrinsic orbital angular momentum probability density $\mathcal{L}_{\text{IOAM}}$ (g--l), and phase $\arg\left[c_{\mathbf{q}}^{(2)}(+\infty)\right]$ (m--r) are shown in the first, second, and third columns, respectively. Each row corresponds to a different number of photons absorbed by the electron-positron pair: from 1 (first row) to 6 (sixth row). The laser frequencies $\omega_0$ are $2m$, $m$, $0.8m$, $0.5m$, $0.4m$, and $0.3m$, respectively. In the colorbar for $\mathcal{L}_{\text{IOAM}}$, positive (negative) values indicate alignment (opposition) of the angular momentum direction with respect to the $q_z$-axis. Other parameters: $\varepsilon_0 = 0.1$, $q_z = 0$, $N = 6$.}
\label{fig:1}
\end{figure*}
\begin{figure*}[!t]\centering
\includegraphics[width=0.33\textwidth]{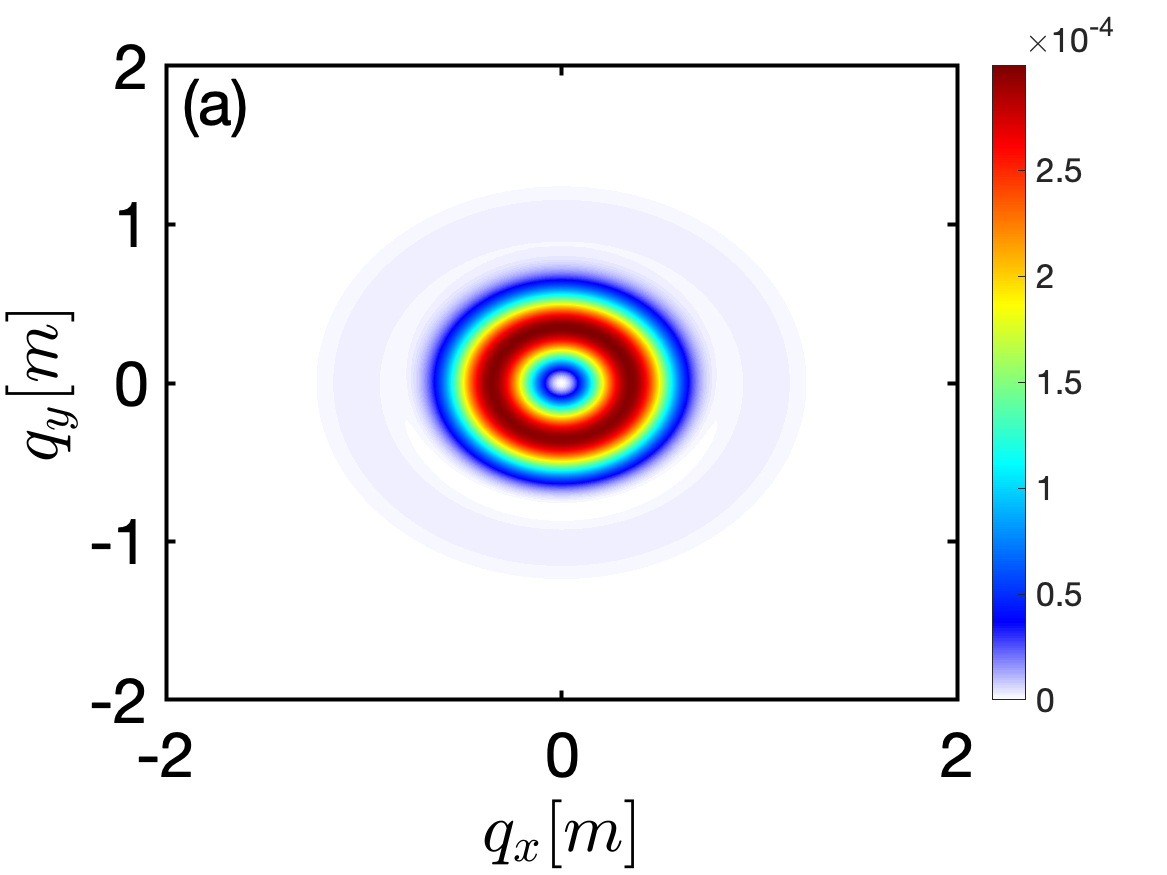}
\includegraphics[width=0.33\textwidth]{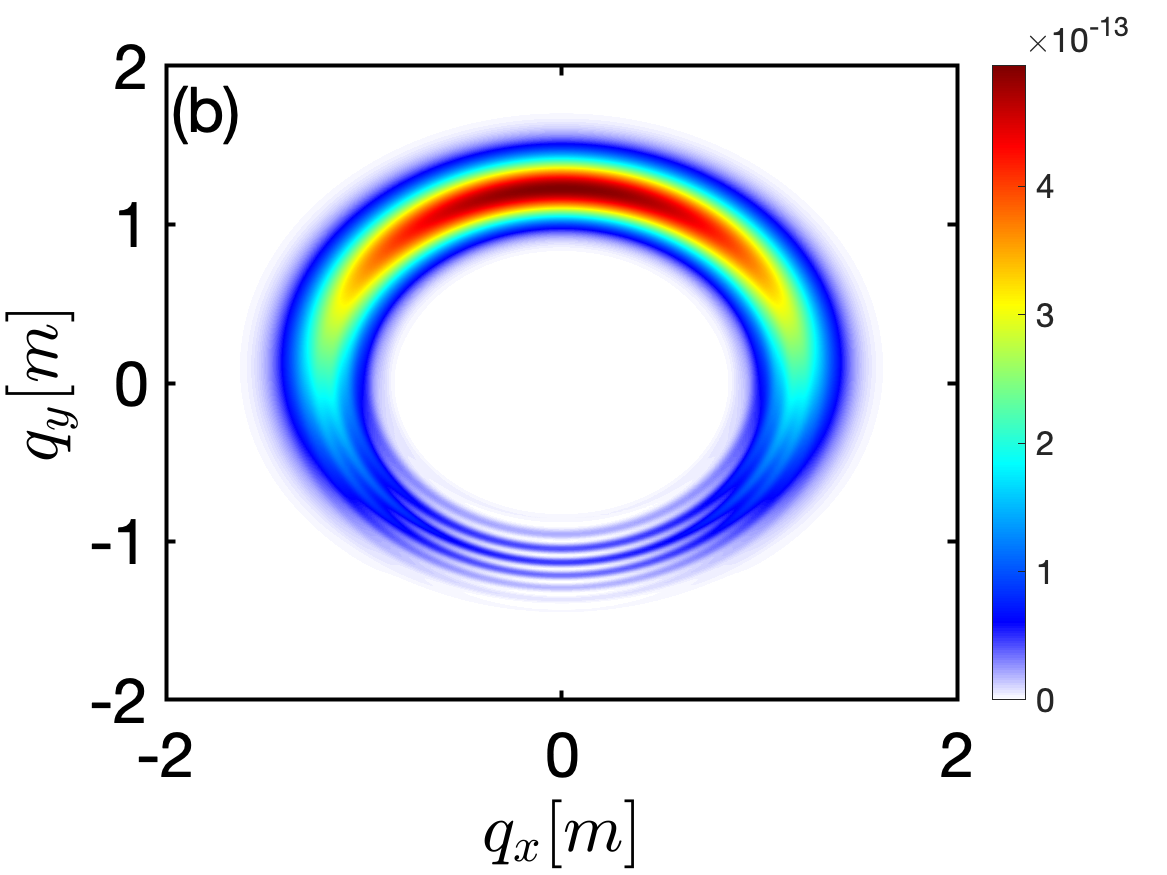}
\includegraphics[width=0.33\textwidth]{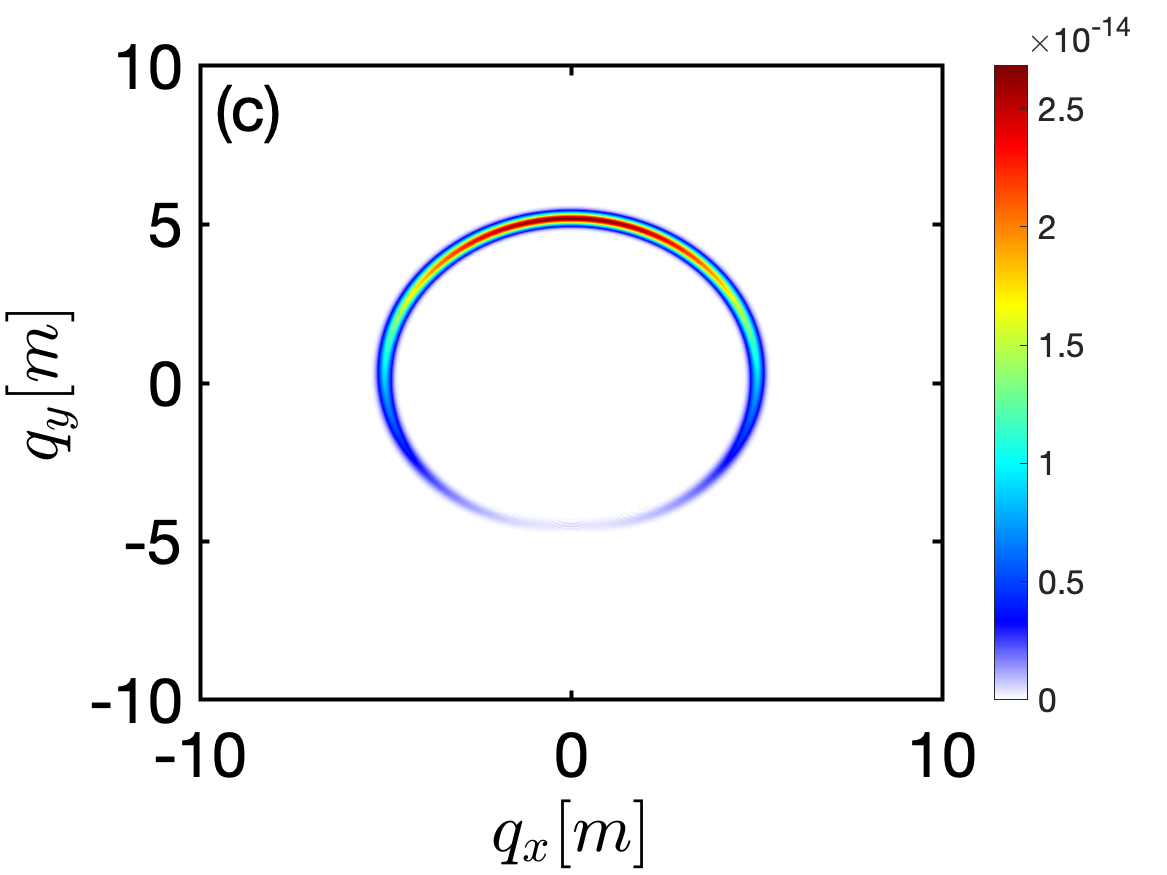}
\includegraphics[width=0.33\textwidth]{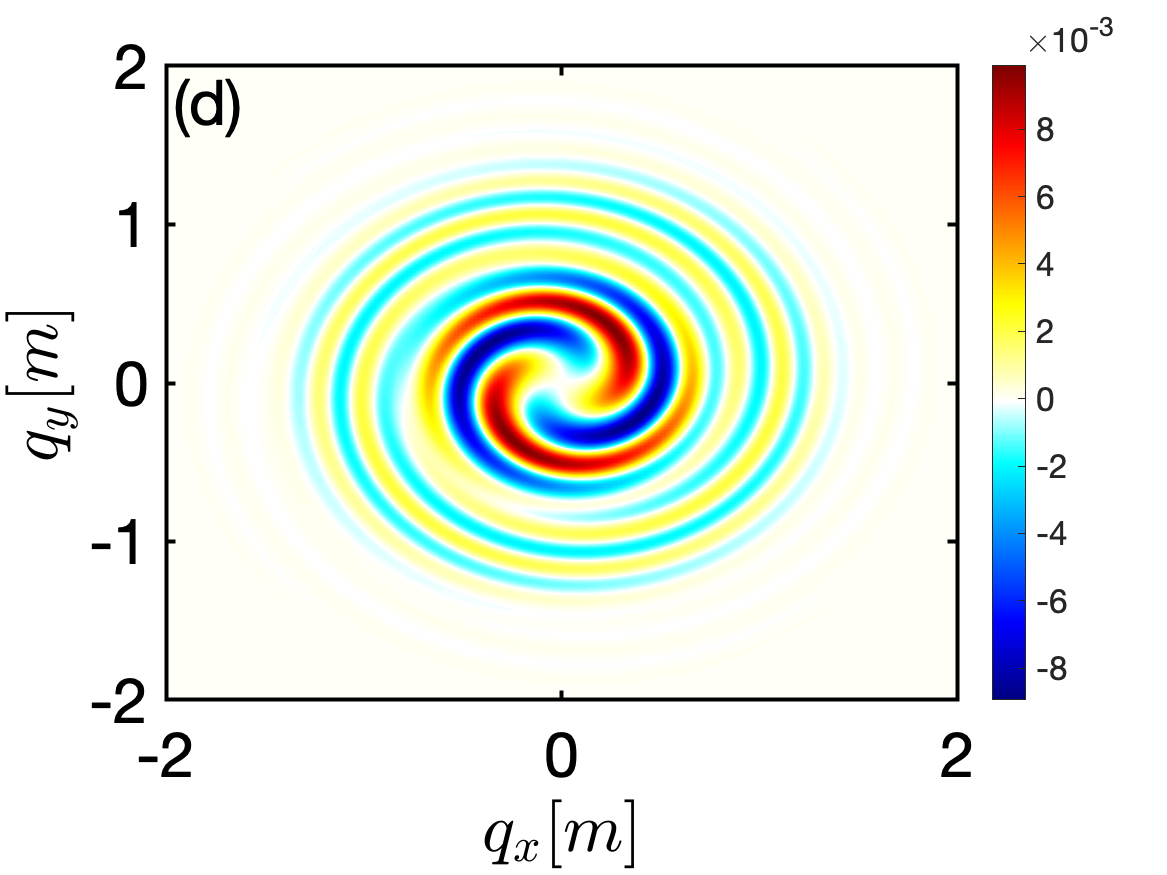}
\includegraphics[width=0.33\textwidth]{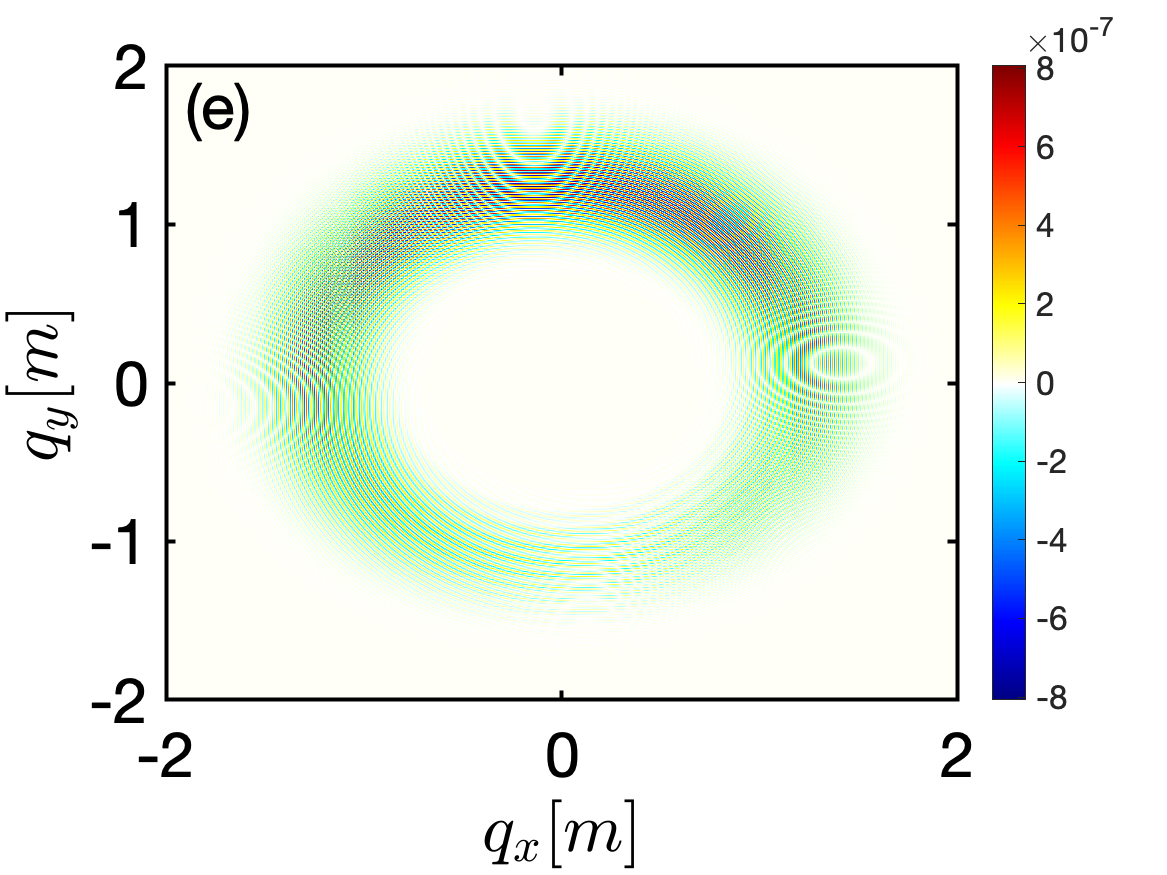}
\includegraphics[width=0.33\textwidth]{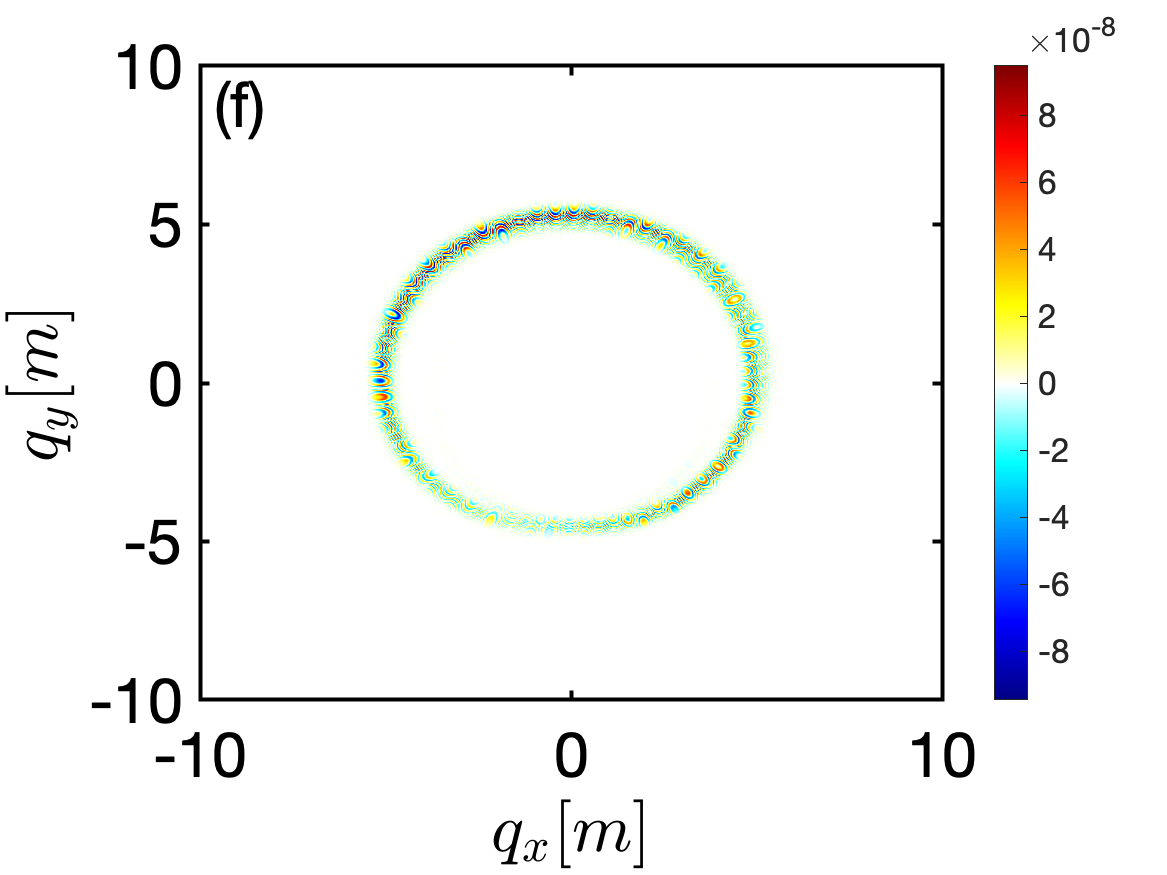}
\includegraphics[width=0.33\textwidth]{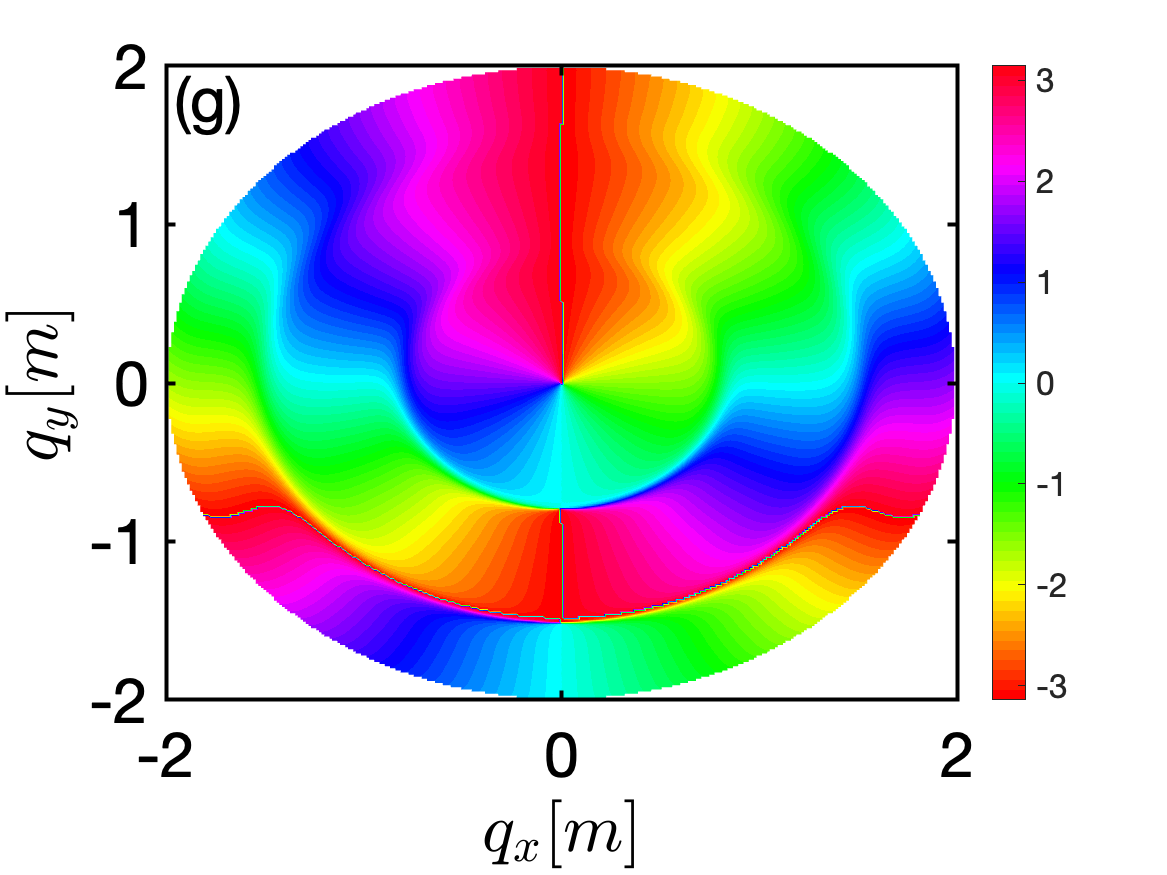}
\includegraphics[width=0.33\textwidth]{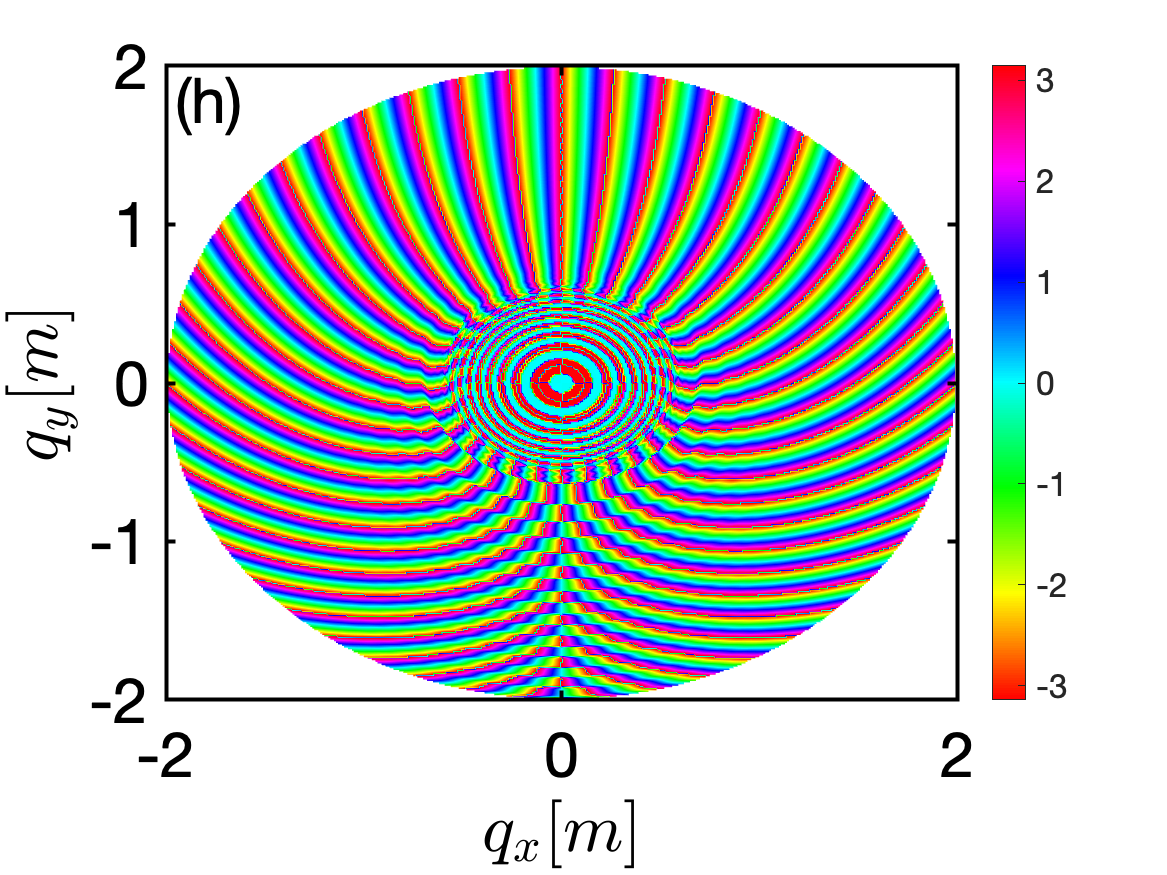}
\includegraphics[width=0.33\textwidth]{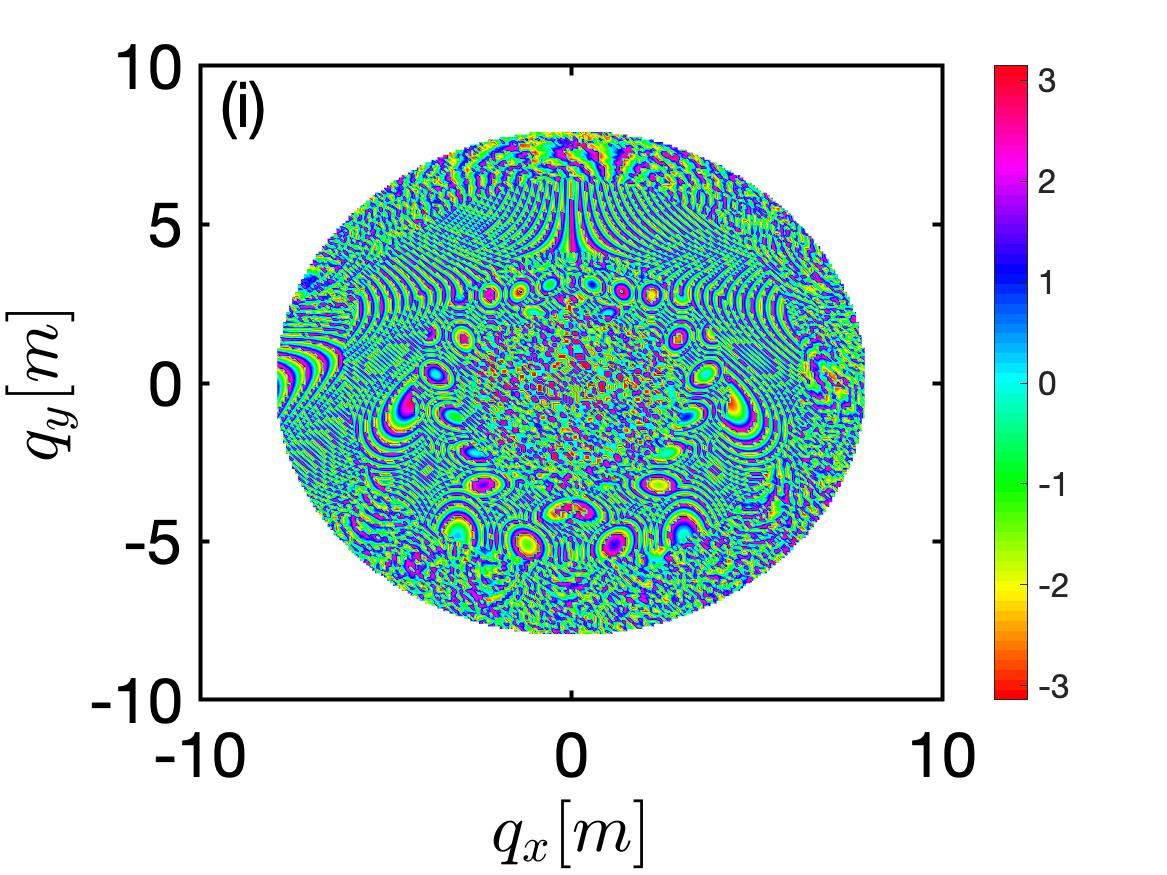}
\caption{First row: electron momentum distribution function $f_{\mathbf{q}}(+\infty)$ (a--c); second row: intrinsic orbital angular momentum probability density $\mathcal{L}_{\text{IOAM}}$ (d--f); third row: phase $\arg\left[c_{\mathbf{q}}^{(2)}(+\infty)\right]$ (g--i). Columns correspond to different pair-production regimes: multiphoton-dominated (a, d, g), mixed-mechanism (b, e, h), and tunneling-dominated (c, f, i). The laser frequencies are $\omega_0 = m$, $0.1m$, and $0.02m$, respectively. In the colorbar for $\mathcal{L}_{\text{IOAM}}$, positive (negative) values indicate alignment (opposition) of the angular momentum with the $q_z$-axis. Other parameters: $\varepsilon_0 = 0.1$, $q_z = 0$, $N = 6$.}
\label{fig:2}
\end{figure*}
Based on the electron momentum distributions shown in Fig.~\ref{fig:1} (a-f), we can predict the magnitude and probability distribution of the intrinsic orbital angular momentum probability density $\mathcal{L}_{\rm IOAM}$ of electron after the multiphoton process (the Keldysh parameter must satisfy the condition $\gamma=m\omega_{0}/e\varepsilon_{0}\gg 1$) using the effective mass $(n\omega_{0}/{2})^2 =q^2_n +m^{2}_{*}$ ($n$ denotes the absorbed photon number), which theoretically should exhibit concentric circles centered at the origin. However, the observed spiral structures in the intrinsic orbital angular momentum probability density $\mathcal{L}_{\rm IOAM}$, shown as in Fig.~\ref{fig:1} (g-l), significantly deviate from the concentric circle pattern. This indicates that simply superimposing the intrinsic angular momentum and its corresponding distribution is not valid. Further analysis reveals that the number of spiral pairs can be estimated using the parameter $n_{s}=\lceil \frac{2m}{\omega_{0}} \rceil$, in which $\lceil \cdot \rceil$ is the ceiling function. The corresponding numbers of spiral pairs $n_{s}$ in Fig.~\ref{fig:1} (a-f) are $1$, $2$, $3$, $4$, $5$, and $6$, respectively. The estimation results are in perfect agreement with the observed spiral structures. Moreover, we observe that the intrinsic orbital angular momentum probability density is entirely confined within the electron momentum distribution.
From Fig.~\ref{fig:1} (m-r), we can also extract the possible values of the IOAM of electrons through phase information, given by the relation $\ell = n_s - s$, where $s = 0, 1, 2, \dots, n_s-1$.  The topological meaning of $\ell$ can be interpreted as the periodicity or the number of repetitions of color variations starting from the origin. Since the electric field we employ is right-hand circularly polarized, the IOAM exhibits a clockwise rotation direction. This result is consistent with the result in Ref.~\cite{Kohlfurst:2022edl}. The phase information can also explain the origin of nodes in the momentum spectrum. For instance, in Fig.\ref{fig:1}(m), no vortex effect is observed near the origin, which is why no node forms at the origin in the electron momentum spectrum shown in Fig.\ref{fig:1}(a). Similarly, in Fig.\ref{fig:1}(n-r), the presence of vortex effects at the origin leads to distinct nodes at the origin in the electron momentum spectra of Fig.\ref{fig:1}(b-f).

Next, by comparing the results of multiphoton dominated, multimechanism dominated ($\gamma=m\omega_{0}/e\varepsilon_{0}\sim 1$), and tunneling dominated ($\gamma=m\omega_{0}/e\varepsilon_{0}\ll 1$) processes, we thoroughly discuss the validity and physical implications of the IOAM, as shown as Fig.~\ref{fig:2} (a-i).
Overall, in the multiphoton-dominated regime, the electron momentum distribution, $\mathcal{L}_{\rm IOAM}$, and phase $\arg\left[c_{\bf{q}}^{(2)}(+\infty)\right]$ exhibit clear features, as shown in Fig.~\ref{fig:2} (a, d, g), requiring no further explanation. However, in the multimechanism-dominated regime, see Fig.~\ref{fig:2}. (b, e, h), both the $\mathcal{L}_{\rm IOAM}$ and $\arg\left[c_{\bf{q}}^{(2)}(+\infty)\right]$ display pronounced spiral structures and interference effects. Notably, the phase information can not accurately estimate the specific values of IOAM, indicating that the effective mass model is insufficient for precisely determining IOAM characteristics. Nevertheless, the $\mathcal{L}_{\rm IOAM}$ still provides information about the total IOAM.
In the tunneling-dominated regime, see Fig.~\ref{fig:2}. (c, f, i), the $\mathcal{L}_{\rm IOAM}$ remains non-zero but manifests solely as pure interference effects. Simultaneously, the phase $\arg\left[c_{\bf{q}}^{(2)}(+\infty)\right]$ exhibits irregular interference patterns, suggesting that the phase information becomes entirely ineffective in the tunneling process. However, according to current theoretical understanding, such phenomena are considered impossible.
From a theoretical perspective, the already operational X-ray free electron laser (XFEL) facilities, such as the Linac Coherent Light Source (LCLS) at SLAC and TESLA at DESY, can in principle achieve near-critical field strengths as large as $E \approx 0.1 E_{cr}$, with corresponding laser frequencies of about $\omega_{0} = 8.3 \, \text{keV} \approx 0.02 \, m$~\cite{Ringwald:2001ib}, corresponding simulation results are shown in Fig.~\ref{fig:2}. (c, f, i). This result suggests that in future pure tunneling experiments, we may observe the IOAM effect. 

\emph{Conclusions.---} 
In this work, we demonstrated that the intrinsic orbital angular momentum (IOAM) originates from axial current density. Without axial current density, the quantum IOAM of particles cannot exist. The spiral or interference characteristics of axial current density are shown to determine the occurrence of nonlinear or tunneling effects within the system. This result is valid for any spacetime-dependent background field, addressing the limitations of the Keldysh's ionization theory and providing a more comprehensive theoretical framework.

Using Wigner function methods, a fermionic generalized two-level model, and Berry phase simulations, we predicted that IOAM effects can persist even in pure quantum tunneling processes. Notably, our findings suggest that it is possible to observe IOAM effects experimentally in future tunneling-dominated regimes, as supported by simulations related to operational XFEL facilities like LCLS and TESLA. These findings open new avenues for exploring strong-field quantum electrodynamics (QED) phenomena and offer valuable insights into the angular momentum properties of quantum systems.

\emph{Acknowledgments.---} We are also grateful to the anonymous referee for critical reading and helpful suggestions to improve the manuscript. This work was supported by the National
Natural Science Foundation of China (NSFC) under Grant Nos. 12447179, 12233002, 12005192, 12273113.  YFH was also supported by the National Key R\&D
Program of China (2021YFA0718500), and by the Xinjiang Tianchi
Program. Cheng-Ming Li was also supported by the Natural Science Foundation of Henan Province of China (No. 242300421375) and by the Project funded by China Postdoctoral Science Foundation (Grant Nos. 2020TQ0287 and 2020M672255). Jin-Jun Geng acknowledges support from the Youth Innovation Promotion Association (2023331).

\end{document}